%% file: SMP-21-013_temp.tex
\pdfoutput=1
\documentclass[11pt,twoside,a4paper,cmspaper,final,collab]{cms-tdr}
\def\svnVersion{550d309}\def\svnDate{2023/09/02}\def\cmsCernNoTag{CERN-EP-2022-270}\def\cmsCernDate{\today}\def\cmsMessage{Published in Physical Review Letters as \href{http://dx.doi.org/10.1103/PhysRevLett.131.091803}{\doi{10.1103/PhysRevLett.131.091803}.}}
\begin{document}\cmsNoteHeader{SMP-21-013}

\newlength\cmsTabSkip\setlength{\cmsTabSkip}{1ex}
\newcommand{\model}{\ensuremath{\,\text{(model)}}\xspace}
\newcommand{\dSh} {\text{dShower}\xspace}
\newcommand{\cuet}{{\textsc{cuetp8m1}}\xspace}
\newcommand{\cpfive}{{\textsc{cp5}}\xspace}
\newcommand{\cuhpp}{{\textsc{cueth}\text{pp}\textsc{s1}}\xspace}
\newcommand{\hw}{{\textsc{herwig7}}\xspace}
\newcommand{\chthree}{{\textsc{ch3}}\xspace}
\newcommand{\sigeff}{\ensuremath{\sigma_{\text {eff}}}\xspace}
\newcommand{\mll}{\ensuremath{m_{\Pell\Pell}}\xspace}
\newcommand{\mZ}{\ensuremath{m_{\PZ}}\xspace}
\newcommand{\pyeight}{{\textsc{pythia8}}\xspace}
\newcommand{\WpmWpm}{\ensuremath{\PW^{\pm}\PW^{\pm}}\xspace}
\newcommand{\dps}{\ensuremath{{\text{DPS}}\,\PW^{\pm}\PW^{\pm}}\xspace}
\newcommand{\lint}{138\,fb$^{-1}$\xspace}
\newcommand{\sigdpsI}{\ensuremath{\sigma^{\WpmWpm}_{\mathrm{DPS}}}\xspace}
\renewcommand{\ie}{{i.e.}\xspace} 
\newcommand{\emu}{\ensuremath{\Pe^{\pm}\PGm^{\pm}}\xspace}
\newcommand{\mumu}{\ensuremath{\PGm^{\pm}\PGm^{\pm}}\xspace}
\newcommand{\Vg}{\ensuremath{\PV\gamma}\xspace}
\newcommand{\ZZ}{\ensuremath{\PZ\PZ}\xspace}
\newcommand{\WZ}{\ensuremath{\PW\PZ}\xspace}
\newcommand{\VVV}{\ensuremath{\PV\PV\PV}\xspace}
\newcommand{\Ptop}{\ensuremath{\PQt}}
\newcommand{\APtop}{\ensuremath{\PAQt}}
\newcommand{\ttV}{\ensuremath{\rm \Ptop\APtop\PV}\xspace}
\newcommand{\ptll}{\ensuremath{\pt^{\Pell\Pell}}{}}
\newcommand{\lone}{\ensuremath{\Pell{1}}\xspace}
\newcommand{\ltwo}{\ensuremath{\Pell{2}}\xspace}
\newcommand{\mttwo}{\ensuremath{\mTii(\Pell\Pell)}\xspace}
\renewcommand{\mT}{\ensuremath{m_{\textrm{T}}}}
\newcommand{\WW}{\ensuremath{\PW\PW}\xspace}
\newcommand{\mupmup}{\ensuremath{\Pgm^{+}\Pgm^{+}}\xspace}
\newcommand{\epmup}{\ensuremath{\Pe^{+}\Pgm^{+}}\xspace}
\newcommand{\ptone}{\ensuremath{\pt^{\lone}}{}}
\newcommand{\pttwo}{\ensuremath{\pt^{\ltwo}}{}}
\newcommand{\etaprod}{\ensuremath{\eta^{\lone}*\eta^{\ltwo}}\xspace}
\newcommand{\etasum}{\ensuremath{\abs{\eta^{\lone} + \eta^{\ltwo}}}\xspace}
\newcommand{\mtone}{\ensuremath{\mT\,(\lone,\ptmiss)}\xspace}
\newcommand{\dphillltwo}{\ensuremath{\abs{\Delta\phi\,(\Pell\Pell,\ltwo)}}{}}
\newcommand{\dphill}{\ensuremath{\abs{\Delta\phi\,(\Pell\Pell)}}{}}
\newcommand{\mtll}{\ensuremath{\mT\,(\Pell\Pell)}{}}
\newcommand{\dphiltwomet}{\ensuremath{\abs{\Delta\phi\,(\ltwo,\ptmiss)}}{}}
\ifthenelse{\boolean{cms@external}}{\providecommand{\suppMaterial}{the supplemental material~\cite{supp}}}{\providecommand{\suppMaterial}{Appendix~\ref{app:suppMat}\nocite{supp}\xspace}}

\cmsNoteHeader{SMP-21-013}

\title{Observation of same-sign \texorpdfstring{$\PW\PW$}{WW} production from double parton scattering in proton-proton collisions at \texorpdfstring{$\sqrt{s} = 13\TeV$}{sqrt(s) = 13 TeV}}

\date{\today}

\abstract{The first observation of the production of \WpmWpm bosons from double parton scattering processes using same-sign electron-muon and dimuon events in proton-proton collisions is reported. The data sample corresponds to an integrated luminosity of 138\fbinv recorded at a center-of-mass energy of 13\TeV using the CMS detector at the CERN LHC. Multivariate discriminants are used to distinguish the signal process from the main backgrounds. A binned maximum likelihood fit is performed to extract the signal cross section. The measured cross section for production of same-sign \PW bosons decaying leptonically is $80.7\pm 11.2\stat^{+9.5}_{-8.6}\syst \pm 12.1\model\unit{fb}$, whereas the measured fiducial cross section is $6.28\pm 0.81\stat \pm 0.69\syst \pm 0.37\model\unit{fb}$. The observed significance of the signal is 6.2 standard deviations above the background-only hypothesis.} 

\hypersetup{%
pdfauthor={CMS Collaboration},%
pdftitle={Observation of same-sign WW from double parton scattering in proton-proton collisions at sqrt(s) = 13 TeV},%
pdfsubject={CMS},%
pdfkeywords={CMS, WW production, DPS}}

\maketitle 

Double parton scattering (DPS) events, in which two hard parton-parton interactions occur in a single proton-proton ($\Pp\Pp$)
collision have been proposed and studied since the advent of the parton model and hadron 
colliders~\cite{Paver:1982yp,Goebel:1979mi,Shelest:1982dg,Sjostrand:1987su,dps2,dps1,Blok:2013bpa,dpsTheory,Bartalini:2017jkk}. The study of
such events sheds light on the internal structure of the colliding protons.
Primarily, the study of DPS processes provides information on the transverse profile of the proton and its energy evolution, 
information that is otherwise not accessible in single parton scattering (SPS) events. In addition, processes with two or 
more hard parton-parton scatterings allow the study of correlations among the partons from the same proton in terms of 
momentum, flavor, color, and spin.

In the simplest theoretical model~\cite{Gaunt}, the two parton-parton interactions 
in DPS can be considered entirely uncorrelated, and 
the expected DPS cross section can be written as the (normalized) product of the SPS cross sections to produce processes A and B independently, as
\begin{linenomath}
\begin{equation}
\label{eq:pocketformula}
    \sigma_{\mathrm{AB}}^{\mathrm{DPS}} = \frac{n}{2} \frac{\sigma_{\mathrm{A}} \sigma_{\mathrm{B}}}{\sigma_{\text{eff}}},
\end{equation}
\end{linenomath}
where $n$ is a combinatorial factor that takes a value of unity if A = B, and two otherwise. The denominator 
\sigeff can be interpreted as being proportional to the average squared transverse distance between the interacting
partons, and it serves as a useful quantity to compare DPS processes in different production modes. 
Its experimental value ranges between 2--10\unit{mb} for gluon-initiated and 10--25\unit{mb} for quark-initiated DPS production processes~\cite{sigeff2,sigeff1,sigeff4,sigeff7,sigeff3,Aad:2014kba,Aaij:2015wpa,Abazov:2015fbl,CMS:2021qsn,sigeff5,sigeff6,cdf:dps,CDF:dps2,LHCb:dps,CMS-FSQ-16-005,atlas:dps2018,CMS:2021lxi,CMS:2021wfx,D0:2014vql}.

As the center-of-mass energy of $\Pp\Pp$ 
collisions increases, so does the density
of sea quarks and gluons, 
which in turn leads to increased DPS contributions to many final states where pairs of 
heavy particles are produced. These additional DPS contributions can limit the precision
of searches for new physics~\cite{CMS:2020cpy} and the accuracy of high-precision standard model analyses~\cite{DelFabbro:1999tf}. 
Therefore, it is important to study DPS processes in different production modes and final states.
Theoretical advancements~\cite{dpsTheory} have improved upon the simple ``geometric'' approach on which Eq.~(\ref{eq:pocketformula}) is based.
The introduction of interparton correlations via double parton distribution 
functions (dPDF)~\cite{Gaunt:2009re,Diehl:2017kgu,Diehl:2018kgr,Gaunt:2018eix}, which include parton splitting 
effects and impact parameter dependence, has led to the first dPDF-based 
Monte Carlo (MC) event generator for DPS events, \dSh~\cite{Cabouat:2019gtm}. 

\begin{figure}[h!]
\centering
\includegraphics[width=0.3\textwidth]{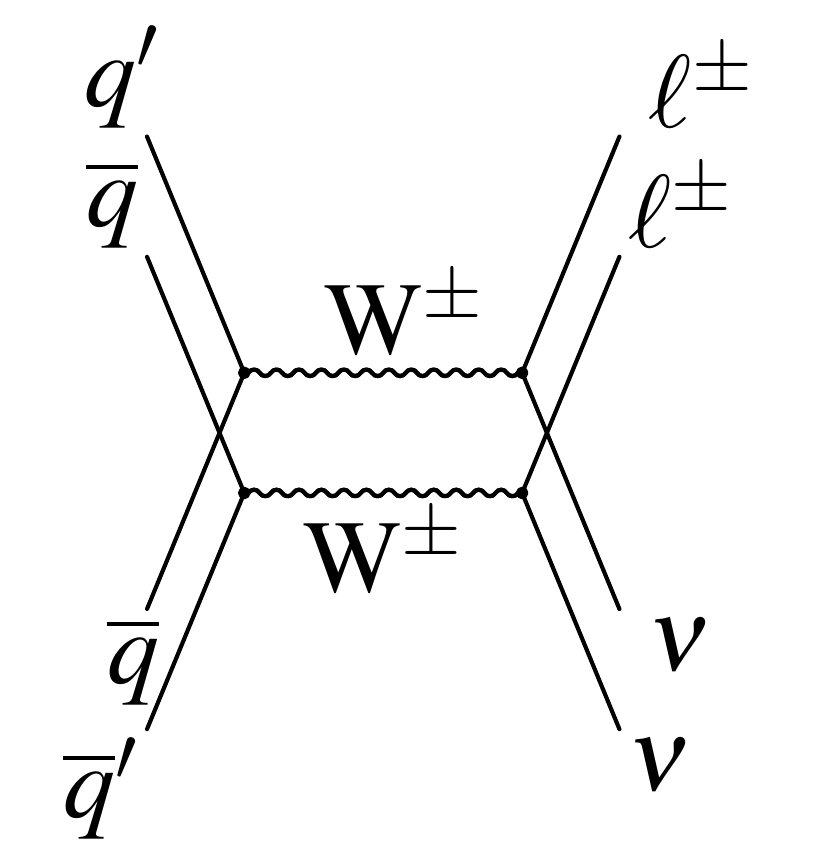}  \\
\includegraphics[width=0.3\textwidth]{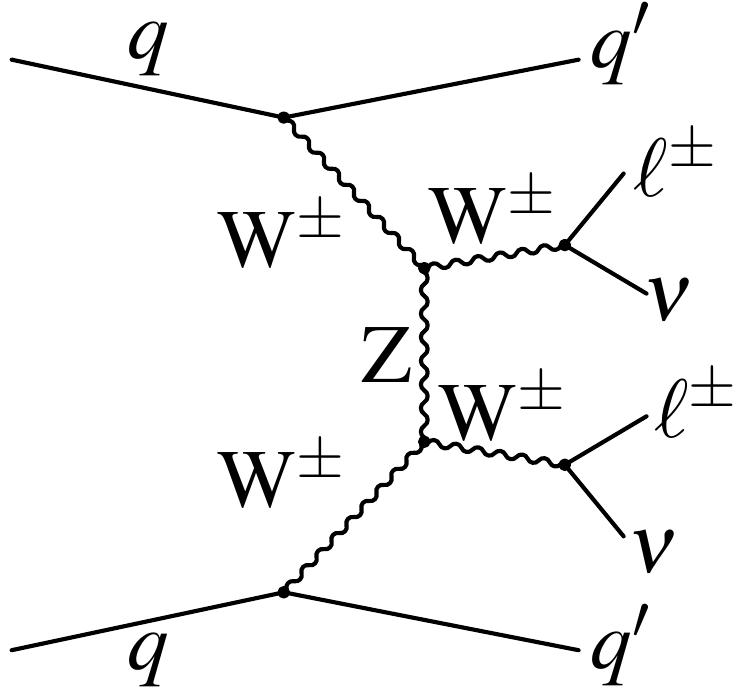}\\
\includegraphics[width=0.3\textwidth]{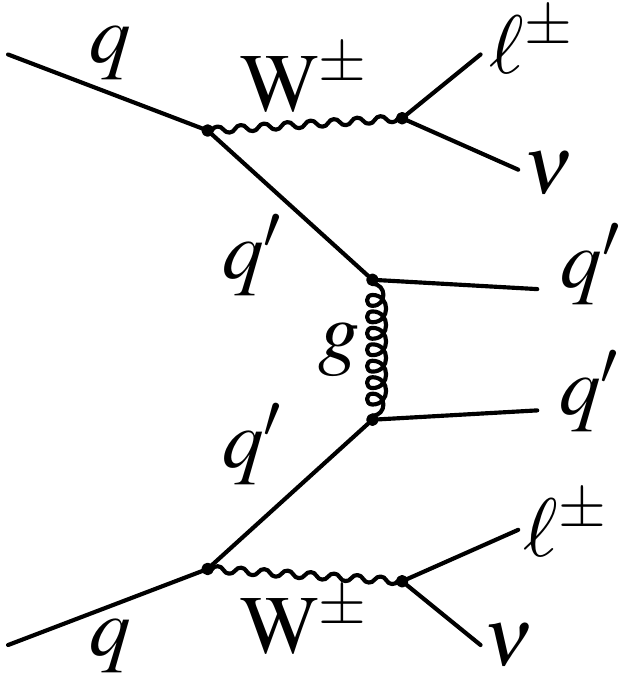}
\caption{Example Feynman diagrams for leptonically decaying $\PW^{\pm}\PW^{\pm}$ bosons produced via DPS (upper) and SPS (middle and lower) processes.}\label{fig:dpsfeyn}
\end{figure}

The production of leptonically decaying same-sign (SS) $\PW$ boson pairs is a 
promising process to study DPS~\cite{Kulesza:1999zh}. Its experimental signature is rather clean and easy to trigger on, and the SPS $\PW^{\pm}\PW^{\pm}$ production is 
highly suppressed because of two additional partons produced in the final state compared with the opposite-sign configuration. 
Figure~\ref{fig:dpsfeyn} illustrates the production of $\PW^{\pm}\PW^{\pm}$
via the DPS and SPS processes at leading order (LO) accuracy in electroweak and perturbative 
quantum chromodynamics (QCD).
The \dps production has not
been observed experimentally, and the existing searches for the process are not statistically significant due to the size 
of the available data samples~\cite{Sirunyan:2019zox,CMS-FSQ-16-005}. 

This Letter presents the first observation of \dps production
using $\Pp\Pp$ collision data at $\sqrt{s} = 13\TeV$. The data sample corresponds to an 
integrated luminosity of \lint collected using the CMS detector  
during the 2016--2018 operation of the LHC.
The $\PW$ bosons are required to decay into final states consisting of two SS leptons ($\Pepm\PGmpm$ or $\PGmpm\PGmpm$), 
including the contributions from leptonic $\tau$ decays. 
As in Refs.~\cite{Sirunyan:2019zox,CMS-FSQ-16-005}, the dielectron final state is not considered 
because of the larger background, but 
the overall analysis strategy has been reoptimized to enhance the signal sensitivity.

The CMS apparatus~\cite{Chatrchyan:2008zzk} is a multipurpose, nearly hermetic detector, 
designed to trigger on~\cite{CMS:2020cmk,Khachatryan:2016bia} and identify electrons, muons, photons, and 
hadrons~\cite{Khachatryan:2015hwa,Sirunyan:2018fpa,CMS:2015myp,CMS:2014pgm}. 
A global ``particle-flow'' (PF) algorithm~\cite{Sirunyan:2017ulk} reconstructs all individual particles
 in an event, combining information from the silicon tracker, the crystal electromagnetic (ECAL),
and the brass-scintillator hadron calorimeters, which all operate within a 3.8\unit{T} superconducting solenoid, 
with data from the gas-ionization muon detectors embedded in the flux-return yoke outside the solenoid.
Details of the event reconstruction used to define the primary vertex (PV) and build leptons, jets, 
hadronically decaying $\tau$ leptons (\tauh), and missing 
transverse momentum (\ptmiss) are provided in Refs.~\cite{Sirunyan:2018pgf,Khachatryan:2016kdb,CMS:2019ctu,Sirunyan:2018pgf,Cacciari:2008gp,FastJet,CMS:2020ebo}. 
Events of interest are selected using a two-tiered trigger system. The first level (L1), composed of custom hardware 
processors, selects events at a rate of around 100\unit{kHz}~\cite{CMS:2020cmk}. The second level, known as the 
high-level trigger, further reduces the event rate to around 1\unit{kHz} before data storage~\cite{Khachatryan:2016bia}.

Charged leptons are required to originate from the primary vertex (PV) to avoid contributions from
additional $\Pp\Pp$ interactions in the same and nearby bunch crossings (pileup).
Electrons are identified using a multivariate analysis (MVA) discriminant that combines 
observables sensitive to the matching of charged-particle tracks in the tracker to the energy deposits in the ECAL
and the amount of bremsstrahlung photons emitted along their trajectory~\cite{Khachatryan:2015hwa,Sirunyan:2020ycc}.
The identification of muons is based on linking track segments reconstructed in the silicon tracker 
to hits in the muon detectors~\cite{Sirunyan:2018fpa}.
Further selection criteria are applied to ensure the correct assignment
of the electric charge of the leptons in the reconstruction~\cite{Khachatryan:2015hwa}. 
Lepton isolation requirements
are imposed following the same approach as in Ref.~\cite{CMS:2020mpn}.
The electron (muon) candidates are required to pass minimal kinematic selection criteria of $\pt > 10\GeV$ 
and $\abs{\eta}<2.5\,(2.4)$ and are referred to as ``loose'' leptons.
The ``tight'' lepton selection used in the analysis employs an MVA discriminant to separate prompt leptons 
coming from the decays of $\PW$ or $\PZ$ (\PV) bosons, or $\Pgt$ leptons and nonprompt leptons~\cite{CMS:2020mpn}. 
The nonprompt leptons originate 
from the decays of light- and heavy-flavor hadrons inside jets produced via strong interactions 
(QCD multijet), hadrons misidentified as leptons, and electrons from photon conversions.
The MVA discriminant is trained using a set of observables related to the lepton kinematics, 
isolation, and identification, as well as variables
relating the lepton to the PV and to the nearest reconstructed PF jet~\cite{CMS:2020mpn}.
Reconstructed jets must not overlap with identified electrons,
muons, or \tauh within $\Delta R = \sqrt{\smash[b]{(\Delta\eta)^2+(\Delta\phi)^2}}<0.4$, where $\phi$ is the azimuthal angle in radians. 
The \textsc{DeepJet} $\PQb$ tagging algorithm~\cite{Bols:2020bkb,CMS-DP-2018-058} is used to identify
jets originating from the hadronization of $\PQb$ quarks.
The chosen working point corresponds to a $\PQb$-tagging efficiency of 84\%  and a mistag rate of 11\% for 
light-flavor quark and gluon jets.

Collision events are collected using a combination of single-lepton and dilepton triggers that require the
presence of one or two isolated leptons above certain \pt thresholds~\cite{Khachatryan:2016bia}.
The resulting trigger efficiency is above 98\% for events passing the subsequent offline selection criteria. 

The dominant background contribution arises from the $\PW\PZ$ process, in which both bosons decay leptonically and one of the
leptons from the $\PZ$ boson decay is either out of
the detector acceptance or does not pass the lepton identification criteria. 
The nonprompt-lepton backgrounds include predominantly $\PW${+}jets
and QCD multijet events, with smaller contributions from \ttbar production.
Other sources of background include $\PW\Pgg^{*}$ events, \Vg events with a photon conversion, and $\PZ\PZ$, 
as well as rare backgrounds, which include SPS $\PW^{\pm}\PW^{\pm}$, \VVV, and \ttV production. 
A minor background contribution stems from Drell--Yan (DY) events when the charge of
one lepton is mismeasured. The charge mismeasurement in dimuon events is negligible~\cite{Sirunyan:2018fpa}, although it is nonzero for electrons~\cite{Khachatryan:2015hwa}.
Hence, the lepton charge misidentification (``charge misid.'') background contributes to the \emu channel only.
The contribution to the signal yield from the production of two overlapping $\PW$ bosons from pileup is negligible. 

The signal process is characterized by the presence of a SS lepton pair with a moderate amount of \ptmiss and little jet activity.
The signal region (SR) selection requires events with exactly one pair of $\Pepm\PGmpm$ or $\PGmpm\PGmpm$ with $\pt>25\,(20)\GeV$ for the leading (subleading)
lepton. To reduce the contribution from \WZ background, events having either a third loosely 
identified lepton with $\pt>10\GeV$ or a $\tauh$ candidate with $\pt>20\GeV$ and $\vert \eta \vert < 2.3$ 
are rejected.
The contributions from nonprompt-lepton backgrounds are reduced by selecting events with $\ptmiss > 15\GeV$ and dilepton invariant mass, $\mll>12\GeV$.

Events are required to have at most one jet with $\pt^{\text{jet}}>30\GeV$ within $\abs{\eta_{\text{jet}}}<2.4$.
Events with a $\PQb$-tagged jet having $\pt^{\PQb\,\text{jet}}>25\GeV$ and $\abs{\eta_{\PQb\,\text{jet}}}<2.4$ are vetoed to reject top quark events. 
For the \emu channel, the dilepton transverse momentum should satisfy $\ptll>20\GeV$ to suppress contributions from \Vg and DY processes.
To increase the signal sensitivity, 
events in the SR are split into four lepton flavor and charge categories: 
$\Pep\PGmp$, $\Pem\PGmm$, $\PGmp\PGmp$, and $\PGmm\PGmm$.

The normalization of the \WZ (\ZZ) background is estimated using $\WZ\to3\Pell\Pgn$ ($\ZZ\to4\Pell$) lepton control regions (CR)
by means of a maximum likelihood (ML) fit to the invariant mass of the three- (four-)lepton system.
The $\WZ\to3\Pell\Pgn$ CR is defined by requiring three tight leptons including at least one opposite-sign same-flavor (OSSF) 
lepton pair with $\abs{\mll -\mZ}<10\GeV$. The invariant mass of the three-lepton system must be ${>}100\GeV$. 
The leading lepton is required to have $\pt > 25\GeV$, and trailing leptons with $\pt > 15\GeV$ are selected.
The \ptmiss threshold is raised to 50\GeV for this CR, and the rest of the selection requirements are taken from the SR.
The $\ZZ\to4\Pell$ CR requires an additional fourth tight lepton with $\pt > 10\GeV$ without any requirement on jets and \ptmiss.
Both the OSSF lepton pairs should have $\mll$ compatible with the \PZ boson mass.

Various MC event generators are used to simulate the signal and background processes.
The signal process is simulated at LO using the \PYTHIA~\cite{Sjostrand:2014zea}, \HERWIG~\cite{Bahr:2008pv}, and 
\dSh~\cite{Cabouat:2019gtm} generators. The \PYTHIA sample, with leptonically-decaying 
$\PW$ bosons, is taken as the nominal signal sample and it predicts
a signal cross section value of $\sigdpsI = 86.4\unit{fb}$ with the tune \cuet~\cite{sigeff7}. 
The tune dependence
of the \pyeight cross section is around 40\%, emphasizing the need for an experimental measurement.
The sample from \HERWIG is used for estimating systematic uncertainties.
The \PYTHIA v8.226\,(8.230) using the NNPDF 2.3~\cite{Ball:2012cx} parton distribution functions (PDFs) with the 
underlying tune \cuet (\cpfive~\cite{CP5tune_CMS}) is used 
for the simulation of the 2016 (2017 and 2018) data-taking periods.
The event generator \HERWIGpp (v2.7)~\cite{Bahr:2008pv,Bellm:2015jjp} using the CTEQ6L1~\cite{Pumplin:2002vw}  PDF set 
with the tune \cuhpp~\cite{sigeff7} 
is used for the 2016 data-taking period, 
whereas \hw (v7.1.4) using the NNPDF 3.1~\cite{Ball:2017nwa} with  PDF set \chthree tune~\cite{CMS:2020dqt} is employed for 2017 and 2018.
The \dSh generator uses dPDF~\cite{Martin:2010db,Diehl:2017kgu} developed using the LO MSTW2008~\cite{Martin:2010db} PDF set.
For a given event generator, neither the underlying event tune nor the PDF sets used to
simulate the signal samples impact the kinematic observables relevant to the analysis.

The \WZ background is simulated at next-to-LO (NLO) in QCD with the \MGvATNLO v2.4.2 (2.6.5) 
generator~\cite{MadGraph5_aMCatNLO} using 
the FxFx jet merging scheme~\cite{Frederix:2012ps} for the 2016 (2017 and 2018) data-taking period.
The \MGvATNLO generator is also used to simulate \Vg, \ttV, and triboson (\VVV) production.
The \POWHEG \textsc{box} (v2.0) code~\cite{Nason:2004rx,POWHEG2,POWHEG3} is used to simulate 
SPS \WpmWpm, \ZZ, and $\PW\Pgg^{*}$ production processes.
The simulated samples of backgrounds for the 2016 data-taking period use the 
NNPDF 3.0~\cite{Ball:2013hta,Ball:2011uy} PDF set whereas the NNPDF 3.1 set 
is used for 2017 and 2018. Background simulations and the \dSh event generator are 
interfaced with \pyeight for modeling of the parton showering, hadronization, and underlying event processes, 
which have similar versions and tunes as the signal process.
The CMS detector response is modeled using \GEANTfour~\cite{Geant}, and the simulated events are reconstructed with the
same algorithms used for the data.
Simulated events are weighted to reproduce the pileup distribution measured in the data. 
The average number of pileup interactions was 23\,(32) in 2016\,(2017 and 2018).
Scale factors are applied to simulated samples to 
correct for differences in the reconstruction and selection of physics objects and in the trigger efficiencies 
between simulation and
data. These scale factors are measured with the ``tag-and-probe'' method using DY events~\cite{Sirunyan:2018shy}.

The nonprompt-lepton background is estimated directly from data using 
the lepton misidentification rate method~\cite{CMS:2020mpn}.
The probability for a loose nonprompt lepton to pass the tight lepton 
selection criteria ($f_{p}$) is estimated using a data control sample dominated by nonprompt leptons and is parameterized as a function of \pt and $\abs{\eta}$.
Events in a sideband of the SR, with at least one lepton failing 
the tight lepton selection criteria, are reweighted 
with $f_{p}$ to  
estimate the nonprompt-lepton background contribution.
A similar approach is used to estimate 
the background contribution from ``charge misidentification'' events by applying the lepton charge misidentification rate to opposite-sign events in data.
The charge misidentification rate is about 0.01\,(0.10)\% for electrons in the barrel\,(endcap) regions.

Due to the topological differences between the dominant $\PW\PZ$ and nonprompt-lepton backgrounds, two separate boosted decision tree (BDT) discriminants~\cite{Hocker:2007ht} are 
trained to distinguish the signal from these background components. Addition of the BDT discriminant trained against the nonprompt-lepton backgrounds improves the signal sensitivity by 13\%.
The BDT training against the \WZ sample is done using its simulated
sample, whereas the training against nonprompt leptons is carried out using a ``tight-loose''
control sample in data.
Kinematic differences between the signal and these backgrounds are explored to define a set of input variables 
for the training of two discriminants, which are optimized based on their discriminating power.
These input variables include:
 transverse momenta of the two leptons;
\ptmiss;
product and absolute sum of the $\eta$ of the two leptons;
separation in $\phi$ between the leptons;
separation in $\phi$ between the subleading lepton and \ptmiss;
separation in $\phi$ between the dilepton system and the subleading lepton;
transverse mass of the two leptons;
transverse mass of the leading lepton and \ptmiss; 
and the ``stransverse mass'' of the dilepton and \ptmiss system~\cite{mt21,mt22}.
The two BDT scores are mapped into a 
two-dimensional (2D) plane in both discriminants. Further, the 2D plane is divided into 13 contiguous regions  
on which the final fit is performed.
This division into regions is performed through optimization of the expected signal
significance.

The integrated luminosities for the 2016, 2017, and 2018 data-taking years have  individual uncertainties of 
1.2--2.5\%~\cite{CMS:2017sdi,CMS:2018elu,CMS:2019jhq}, while the overall uncertainty for the 2016--2018 period is 1.6\%. 
The simulation of pileup events assumes an inelastic $\Pp\Pp$ cross section of 69.2\unit{mb}, 
with an associated uncertainty of 5\%~\cite{CMS:2018mlc}, which impacts the expected signal and background yields by 1\%.
The uncertainties in data to simulation scale factors corresponding to 
the lepton trigger, reconstruction, and identification result in an
uncertainty of 3.3\,(2.8)\% for the \emu\,(\mumu) final states. 
This also includes an uncertainty in the scale factors applied to account for the
L1 trigger inefficiency observed in the $\abs{\eta} > 2.0$ region for the 2016 and 2017 data-taking periods~\cite{CMS:2020cmk}.
The jet energy scale uncertainties affect the expected event yields by 3\%.
Uncertainties in the \ptmiss are 
calculated by varying the momenta of unclustered particles that are identified neither 
as a jet nor as a lepton and affect the expected event yields by ${\approx}2\%$.
The uncertainty in the \PQb tagging efficiency has an effect of ${\approx}1.8\%$ on the expected event yields.

For the $\Pepm\PGmpm$\,($\PGmpm\PGmpm$) channels, 
a normalization uncertainty of 30\,(25)\% in the nonprompt-lepton background 
 accounts for the observed variations
in the background estimation method when applied to simulated samples.
The dependence of $f_{p}$ on the kinematics of the event sample in which it is measured is included as a shape uncertainty~\cite{CMS:2020mpn}.
A normalization uncertainty of 20\% is applied to the ``charge misid.'' background, 
covering the differences in the measurement of the
charge misidentification rate in data and simulation. 
The data-to-MC normalization factors of the $\Vg$ background, measured using a 
dedicated CR~\cite{CMS:2021icx}, are close to unity.

A 50\% normalization uncertainty is applied to all other small simulated backgrounds, accounting for the
theoretical uncertainties in the predicted cross sections.
The theoretical uncertainties associated with the choice of the renormalization and factorization scales and 
the variations in the PDFs and the strong coupling constant~\cite{Ball:2014uwa,Butterworth:2015oua} 
affect the simulated background yields by ${\approx}1\%$. 
The effect of variation in PDF sets is negligible for the signal simulations.
Any residual model dependence of the signal process is estimated by allowing the shape of the
final BDT discriminant to vary between the \PYTHIA and \HERWIG simulations and the resulting effect is small. 
The statistical uncertainties due to the limited size of the MC samples are treated according to
the Barlow--Beeston--lite method~\cite{Barlow:1993dm,Conway:2011in}.

\begin{figure}[h!]
\centering
\includegraphics[height=0.265\textheight]{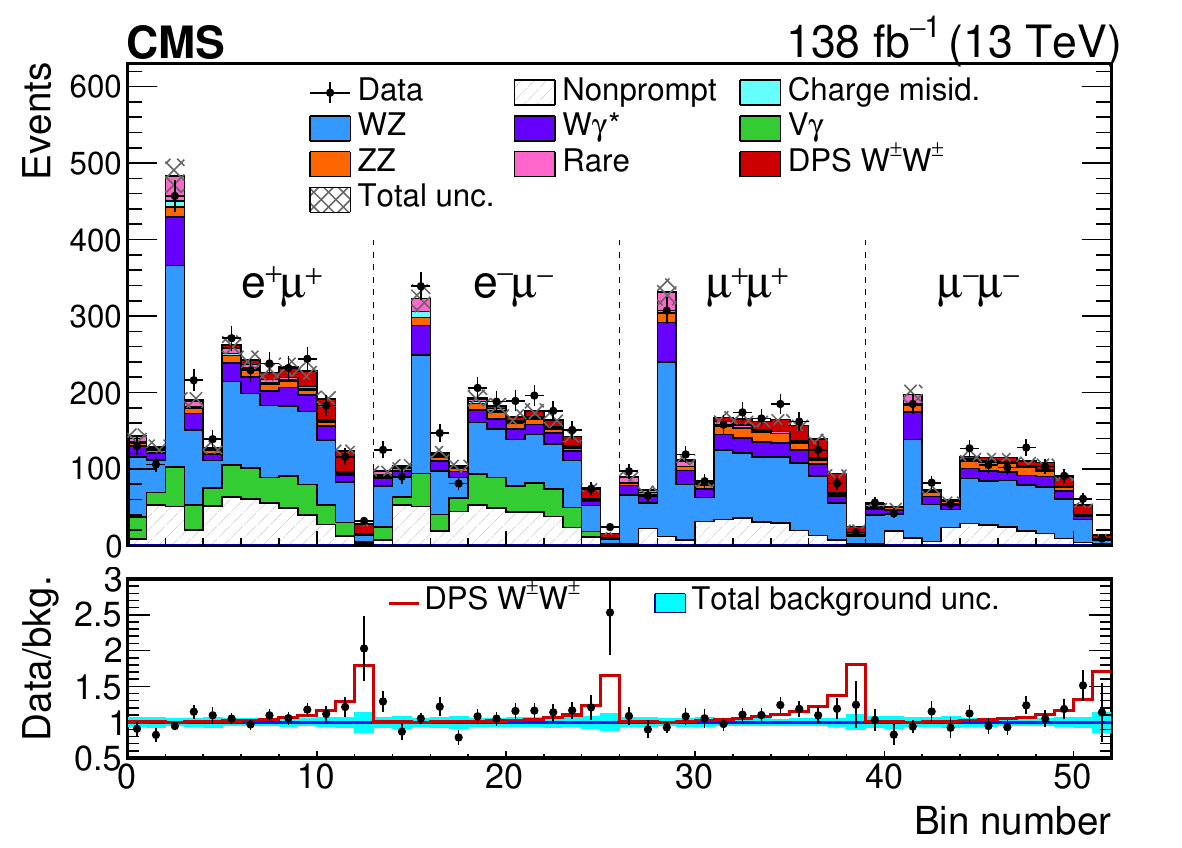}
\caption{Postfit distribution of the final BDT discriminant output for the 
four lepton flavor and sign categories.
The SPS $\PW^{\pm}\PW^{\pm}$, \ttV, and \VVV contributions are grouped as the ``Rare'' background.
The total postfit uncertainty in the signal and background predictions is shown as the hatched band.
The bottom panels show the ratio of data to the sum of all background contributions as the black
data points along with the extracted signal shown by the red line.
The vertical error bars on the data points represent the statistical uncertainty of the data.}\label{fig:results}
\end{figure}

The yield of the \dps process in the SR is obtained by performing a binned ML fit simultaneously in the SR and in the 
two CRs~\cite{Cowan:2010js,CMS-NOTE-2011-005,Khachatryan:2014jba}, \ie, the normalization factors for the
\WZ and \ZZ backgrounds are included as free parameters in the fit together with the signal process.
The fit is performed after combining all the background and signal processes in the 
aforementioned lepton flavor and charge categories, 
resulting in four independent distributions of the final BDT discriminant per data-taking year.
An excess of events with respect to the background-only hypothesis is observed, which is quantified by calculating the 
p-value using a profile likelihood ratio test statistic~\cite{Cowan:2010js}.
Figure~\ref{fig:results} shows the BDT discriminant distribution after 
the ML fit (postfit) for the four lepton flavor and sign categories.
The contributions from different backgrounds and the signal processes 
are stacked on top of each other and the associated postfit uncertainties are also shown. 
The distributions of the kinematic variables used to train the BDT discriminants along with the two BDT discriminants 
are shown in \suppMaterial{}.

The measured value of \sigdpsI is $80.7\pm 11.2\stat^{+9.5}_{-8.6}\syst \pm 12.1\model\unit{fb}$, where the 
model uncertainty
accounts for the difference in cross sections obtained in the experimental 
acceptance region with the \PYTHIA and \HERWIG simulations.
The observed statistical significance of the signal is 6.2 standard deviations above the background-only hypothesis. 
Separate fits to the \emu and \mumu channels indicate that the two measurements agree
within 1.7 standard deviations. 
The \dps production cross section is also measured in a fiducial volume, 
defined using two generator-level SS ``dressed'' leptons (\emu or \mumu) from $\PW$ boson decays excluding the 
events with leptonically decaying $\tau$ leptons.
The leptons are dressed by adding the momenta of generator-level photons 
within a cone of $\Delta R(\Pell,\gamma) < 0.1$ to their momenta, and are required to pass 
kinematic requirements on the \pt, $\eta$, $\mll$, and $\ptll$ variables from the SR selection. 
The measured fiducial cross section is $6.28\pm 0.81\stat \pm 0.69\syst \pm 0.37\model\unit{fb}$, where the model uncertainty 
represents the observed difference in reconstruction efficiencies within the fiducial region obtained using the \PYTHIA and \HERWIG simulations.
The measured value of the 
inclusive (fiducial) cross section is in agreement with the predicted value of 86.4\,(6.74)\unit{fb} by \pyeight 
with the tune \cuet and \dSh. 
A value of $\sigma_{\text{eff}}$ is extracted from Eq.~(\ref{eq:pocketformula}), using the measured \sigdpsI value and the 
next-to-NLO prediction for the single \PWp\,(\PWm) production cross section including leptonic decays of $35.4\pm 1.4\,(26.0 \pm 1.0)\unit{nb}$~\cite{Anastasiou:2003ds,fewz2p1}.
This procedure results in a value of $12.2^{+ 2.9}_{- 2.2}\unit{mb}$, consistent with previous measurements of this quantity from
final states with vector bosons~\cite{atlas:dps2018,CMS:2021qsn}.
Tabulated results are provided in HEPData~\cite{HEPData}.

In summary, the first observation of \WpmWpm production from double parton scattering processes in proton-proton
collisions at $\sqrt{s} = 13\TeV$ has been reported. The analyzed data set corresponds to an integrated
luminosity of \lint, collected in 2016--2018 using the CMS detector at the LHC.
Events are selected by requiring same-sign electron-muon or dimuon 
pairs with moderate missing transverse momentum and low jet multiplicity.
Boosted decision trees are used to discriminate between
the signal and the dominant background processes.
A fiducial cross section of $6.28\pm 0.81\stat \pm 0.69\syst \pm 0.37\model\unit{fb}$ is extracted, and
an inclusive cross section of $80.7\pm 11.2\stat^{+9.5}_{-8.6}\syst \pm 12.1\model\unit{fb}$
is measured.  This corresponds to an observed significance of the signal above the background-only hypothesis
of 6.2 standard deviations.
A value of the DPS effective cross section, characterizing the transverse distribution of partons 
in the proton, $\sigma_{\text{eff}}=12.2^{+ 2.9}_{- 2.2}\unit{mb}$ is extracted.

\begin{acknowledgments}
We congratulate our colleagues in the CERN accelerator departments for the excellent performance of the LHC and thank the technical and administrative staffs at CERN and at other CMS institutes for their contributions to the success of the CMS effort. In addition, we gratefully acknowledge the computing centers and personnel of the Worldwide LHC Computing Grid and other centers for delivering so effectively the computing infrastructure essential to our analyses. Finally, we acknowledge the enduring support for the construction and operation of the LHC, the CMS detector, and the supporting computing infrastructure provided by the following funding agencies: BMBWF and FWF (Austria); FNRS and FWO (Belgium); CNPq, CAPES, FAPERJ, FAPERGS, and FAPESP (Brazil); MES and BNSF (Bulgaria); CERN; CAS, MoST, and NSFC (China); MINCIENCIAS (Colombia); MSES and CSF (Croatia); RIF (Cyprus); SENESCYT (Ecuador); MoER, ERC PUT and ERDF (Estonia); Academy of Finland, MEC, and HIP (Finland); CEA and CNRS/IN2P3 (France); BMBF, DFG, and HGF (Germany); GSRI (Greece); NKFIH (Hungary); DAE and DST (India); IPM (Iran); SFI (Ireland); INFN (Italy); MSIP and NRF (Republic of Korea); MES (Latvia); LAS (Lithuania); MOE and UM (Malaysia); BUAP, CINVESTAV, CONACYT, LNS, SEP, and UASLP-FAI (Mexico); MOS (Montenegro); MBIE (New Zealand); PAEC (Pakistan); MES and NSC (Poland); FCT (Portugal);  MESTD (Serbia); MCIN/AEI and PCTI (Spain); MOSTR (Sri Lanka); Swiss Funding Agencies (Switzerland); MST (Taipei); MHESI and NSTDA (Thailand); TUBITAK and TENMAK (Turkey); NASU (Ukraine); STFC (United Kingdom); DOE and NSF (USA).
\end{acknowledgments}

\bibliography{auto_generated}

\ifthenelse{\boolean{cms@external}}{}{
\clearpage

\appendix
\numberwithin{figure}{section}
\section{Supplemental material\label{app:suppMat}}
\input{supplemental_material}

}

\cleardoublepage \section{The CMS Collaboration \label{app:collab}}\begin{sloppypar}\hyphenpenalty=5000\widowpenalty=500\clubpenalty=5000\input{SMP-21-013-public-authorlist.tex}\end{sloppypar}
%%% END EDITABLE REGION %%%
% skeleton_end
\end{document}

%% file: supplemental_material.tex
\begin{figure*}[htbp!]
\centering
\includegraphics[width=0.25\textwidth]{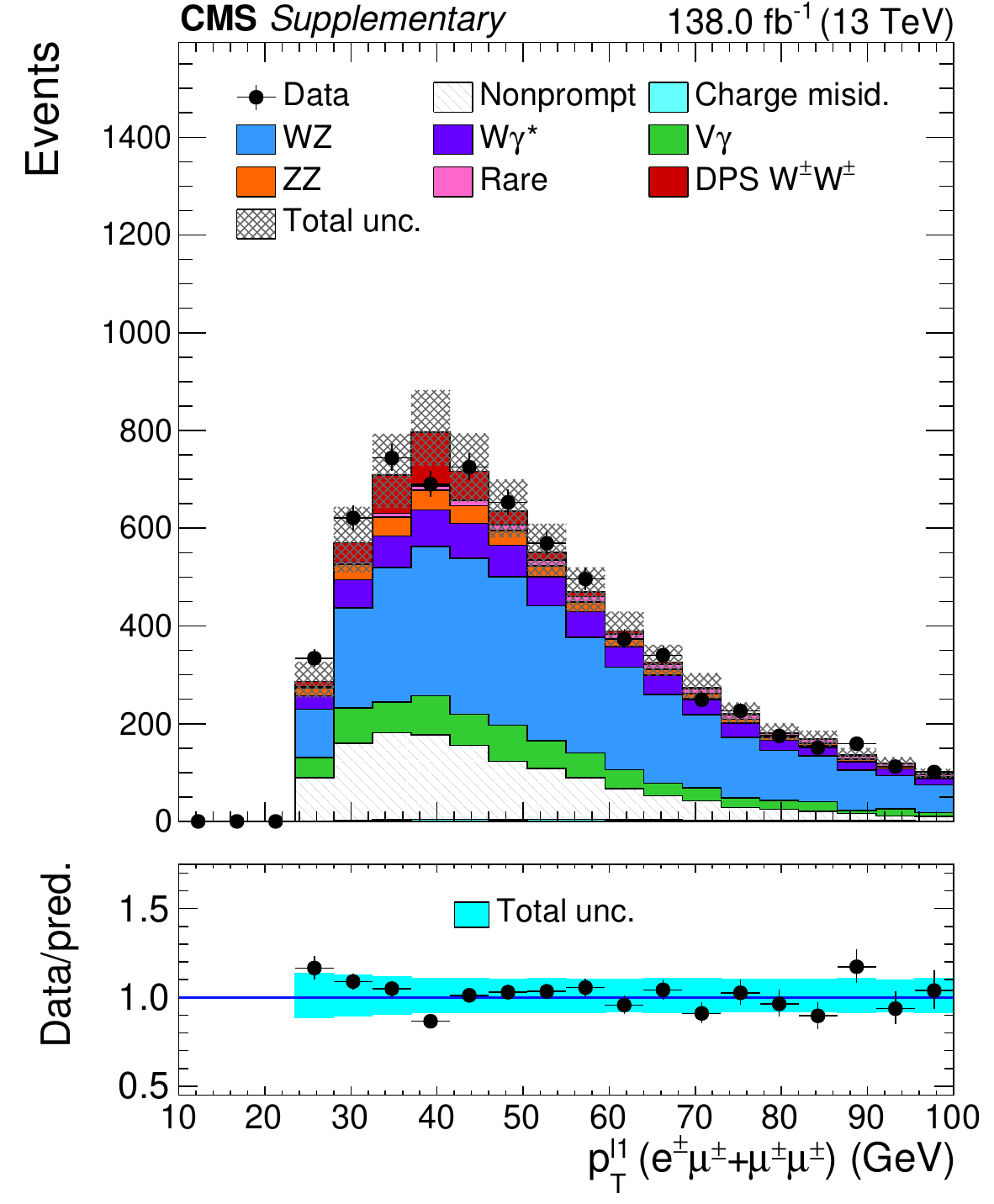}%pt1
\includegraphics[width=0.25\textwidth]{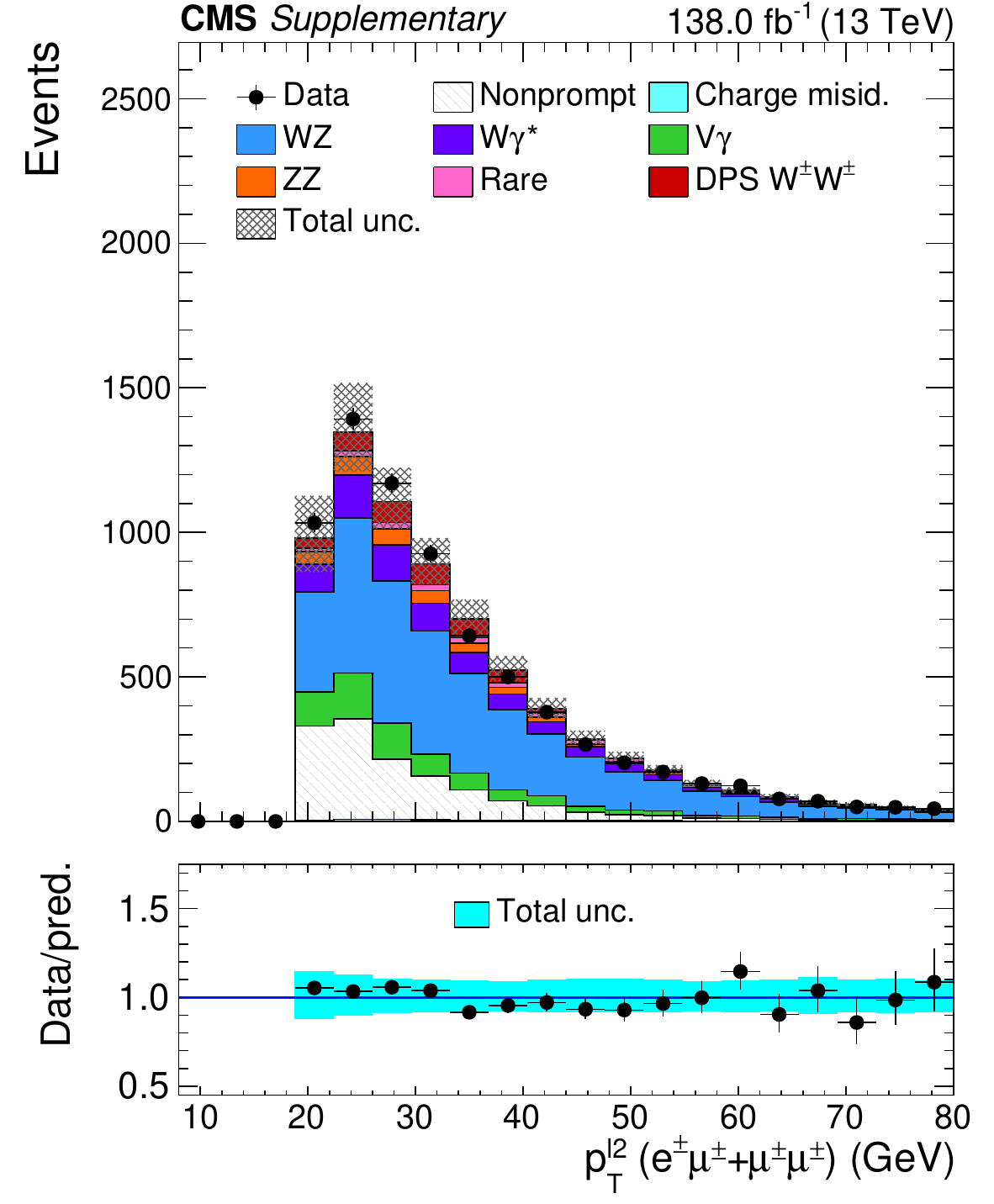}%pt2
\includegraphics[width=0.25\textwidth]{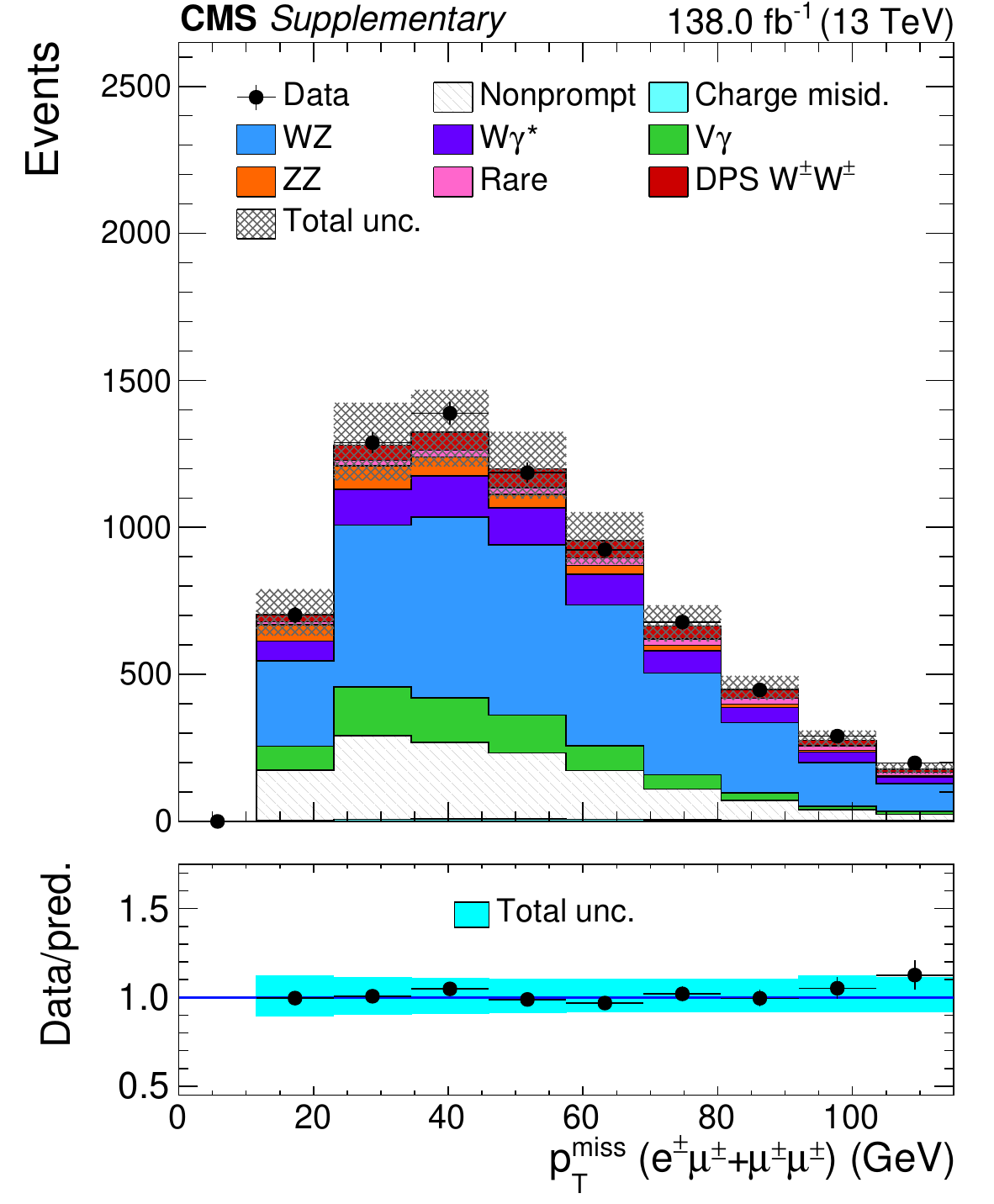}%ptmiss
\includegraphics[width=0.25\textwidth]{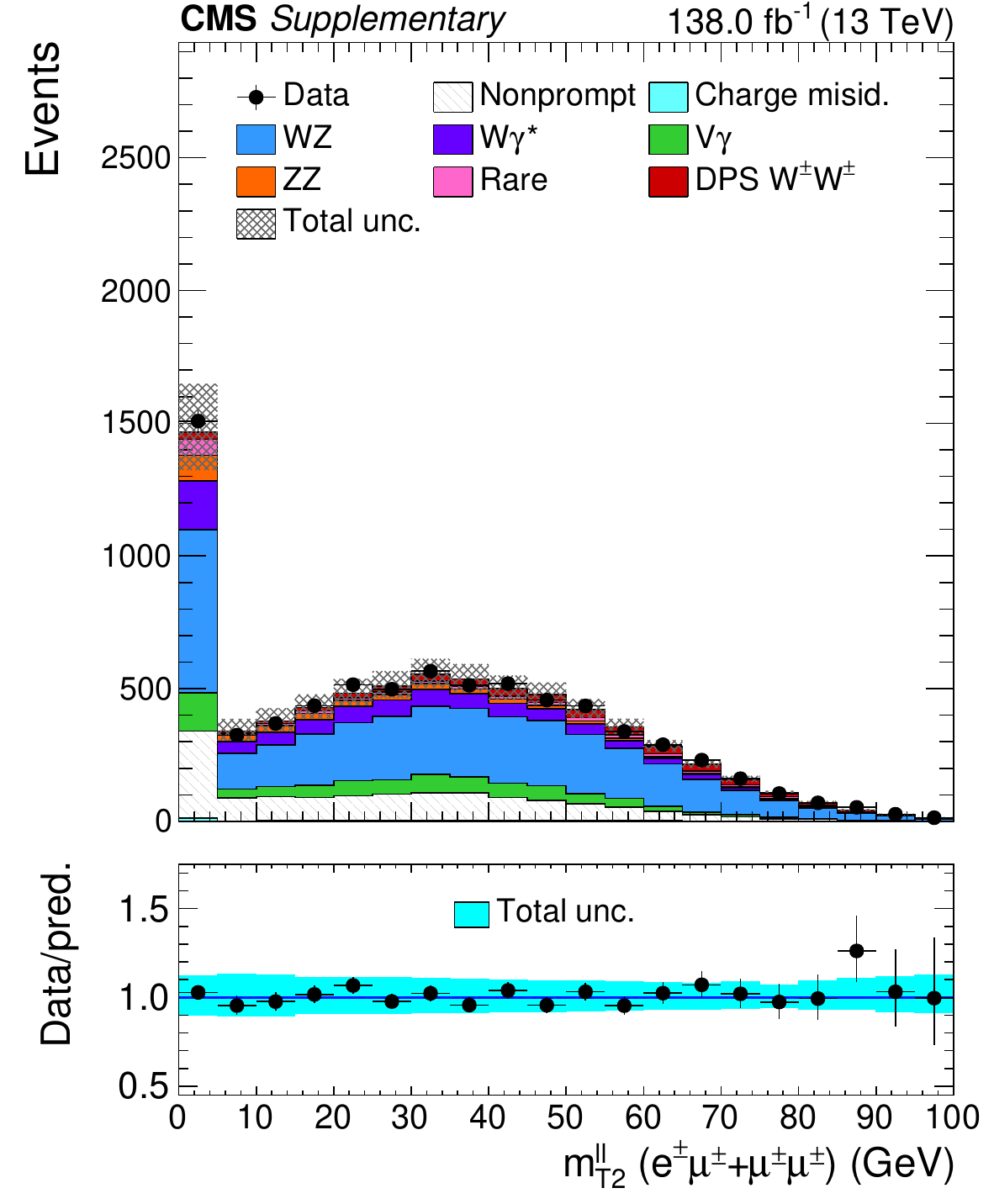}\\ %mT2
\includegraphics[width=0.25\textwidth]{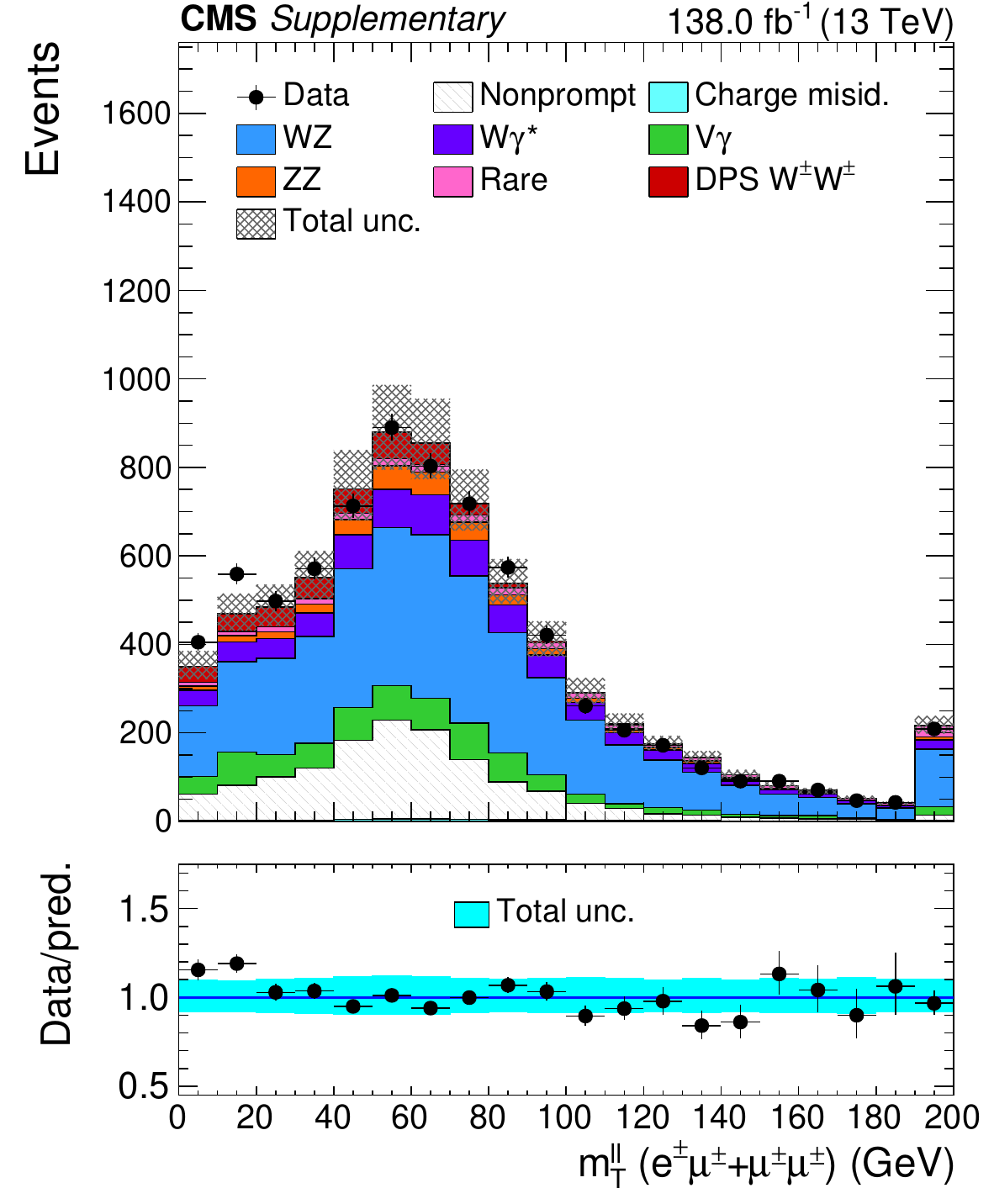}%mT
\includegraphics[width=0.25\textwidth]{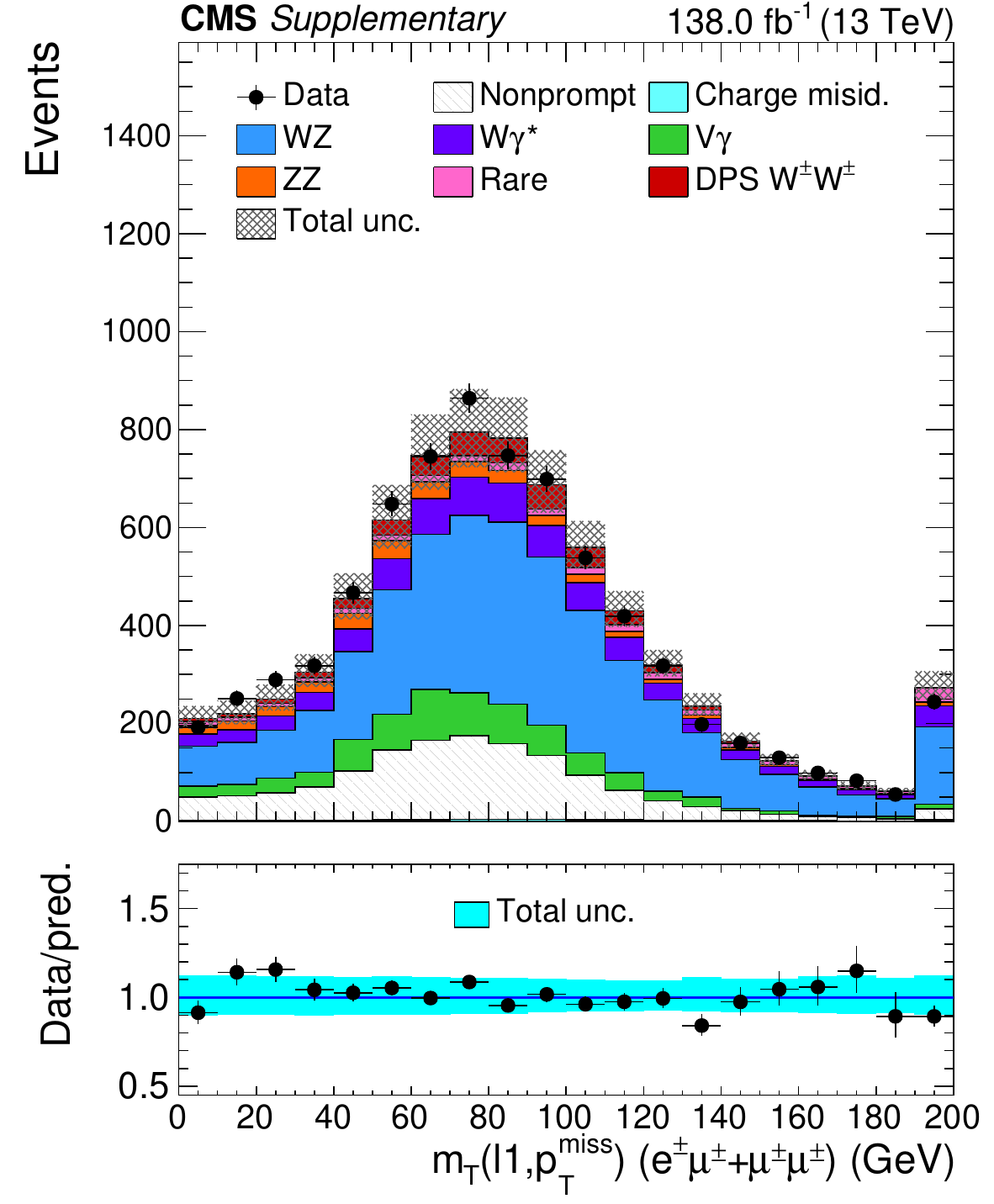}%mT1
\includegraphics[width=0.25\textwidth]{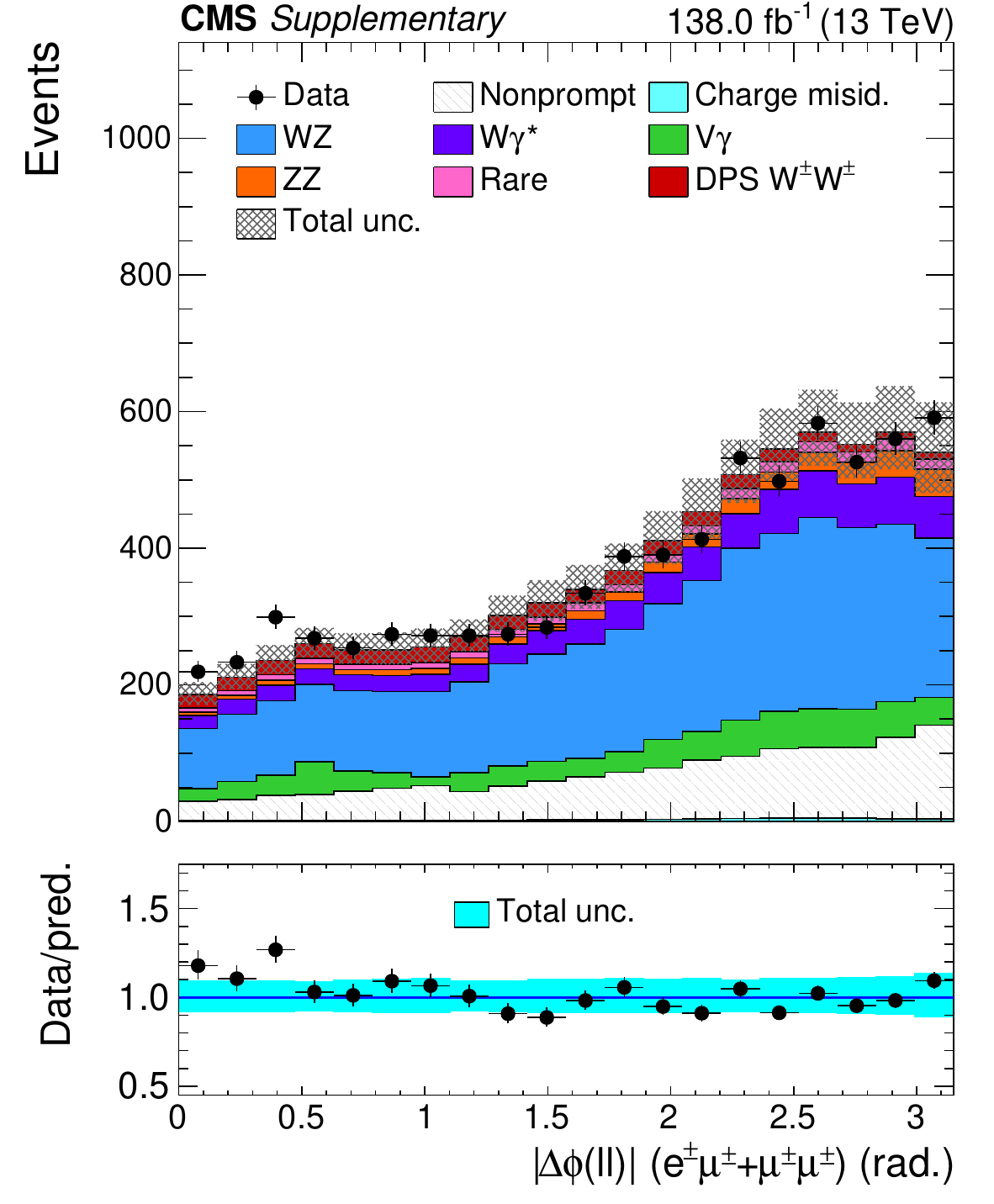}%dphill
\includegraphics[width=0.25\textwidth]{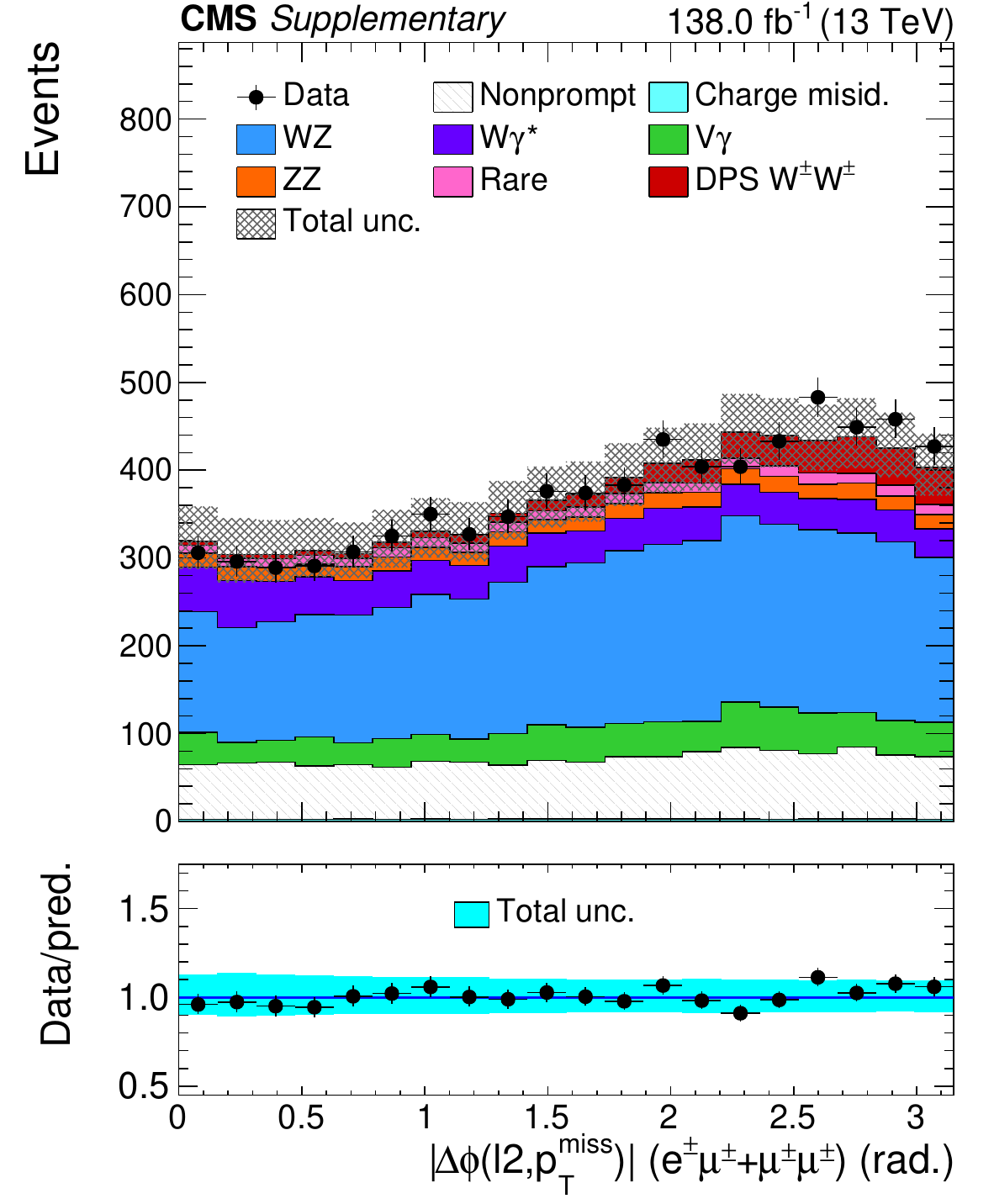}\\ %dphil2met
\includegraphics[width=0.25\textwidth]{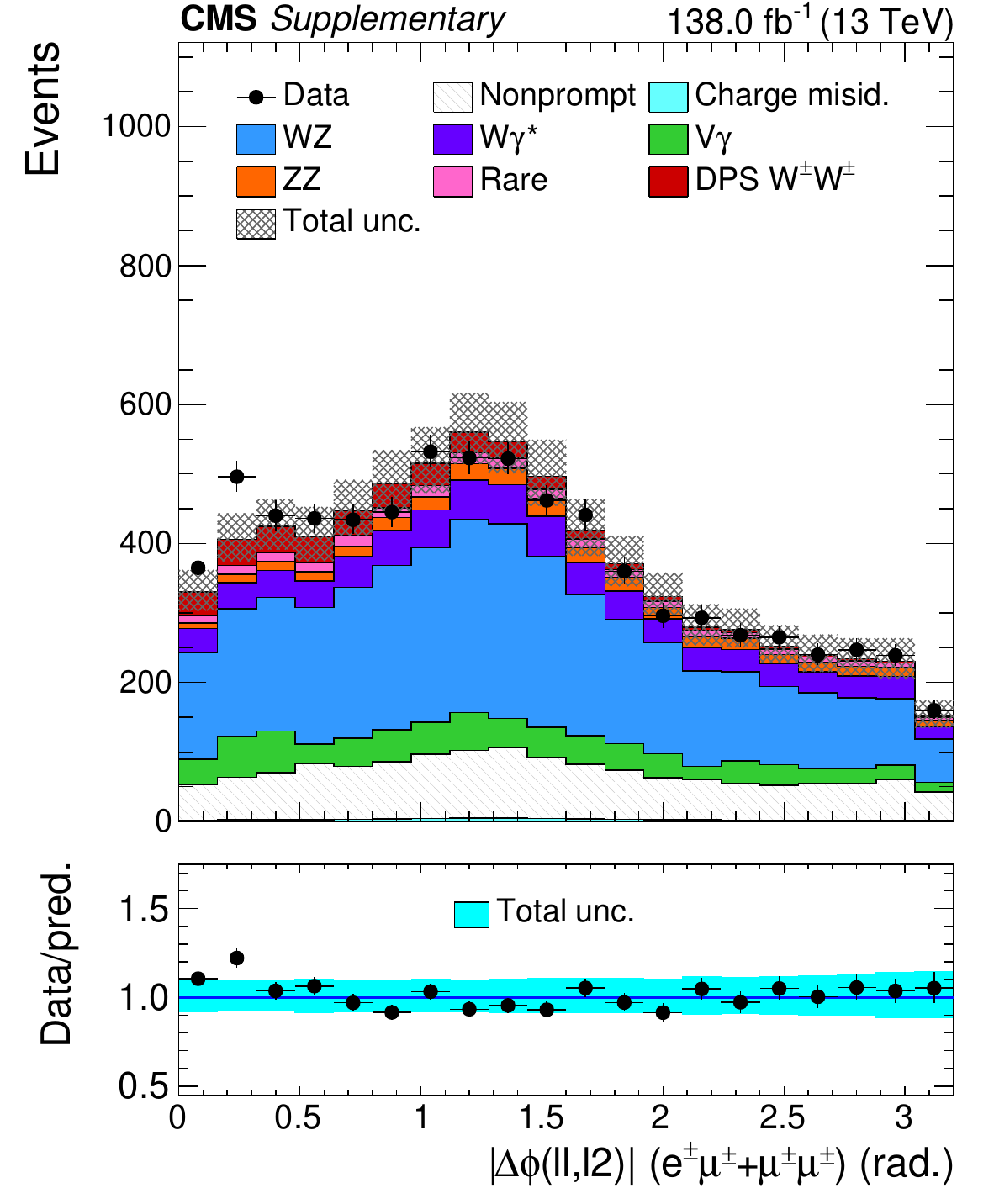}%dphilll2
\includegraphics[width=0.25\textwidth]{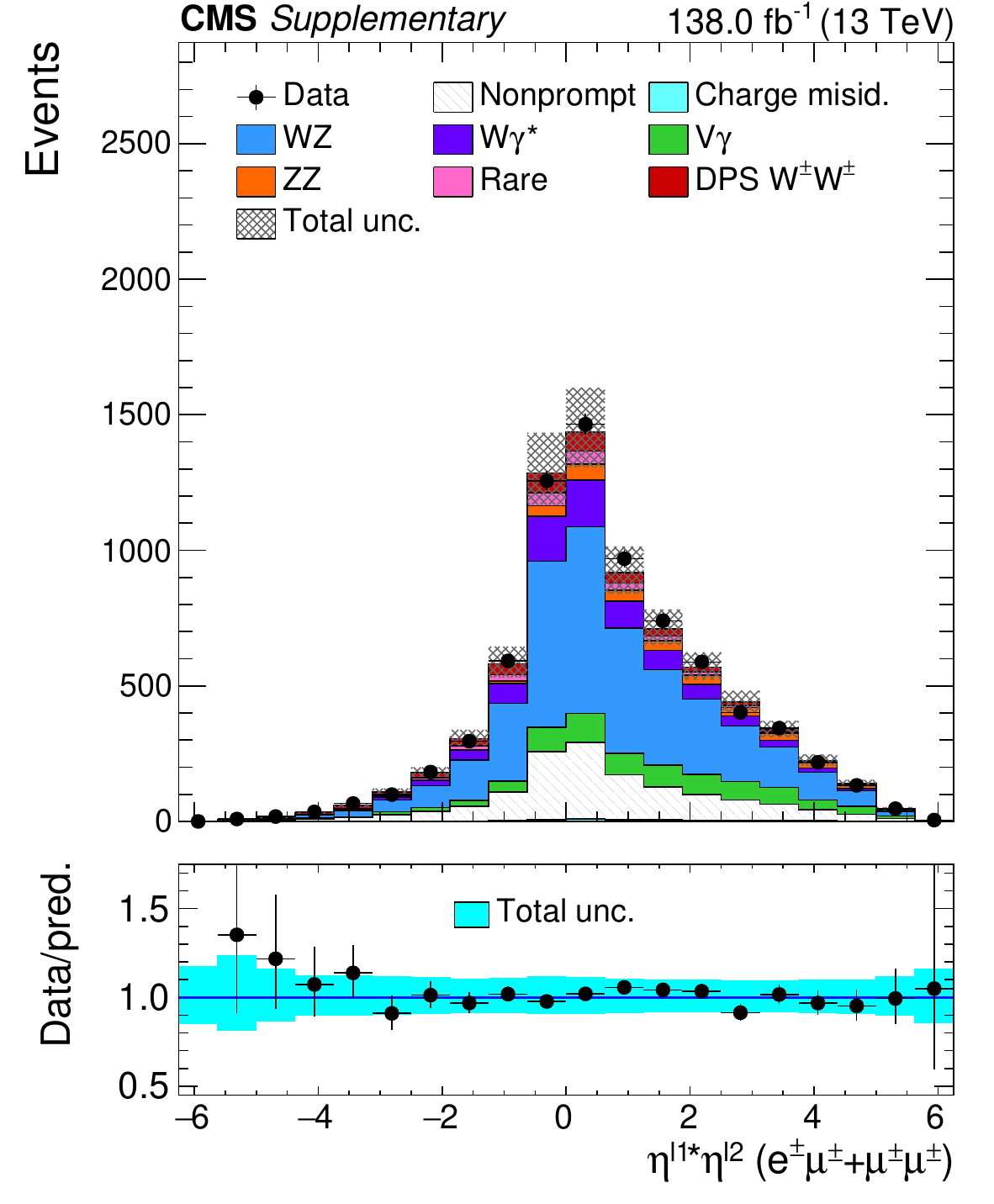}%etaP
\includegraphics[width=0.25\textwidth]{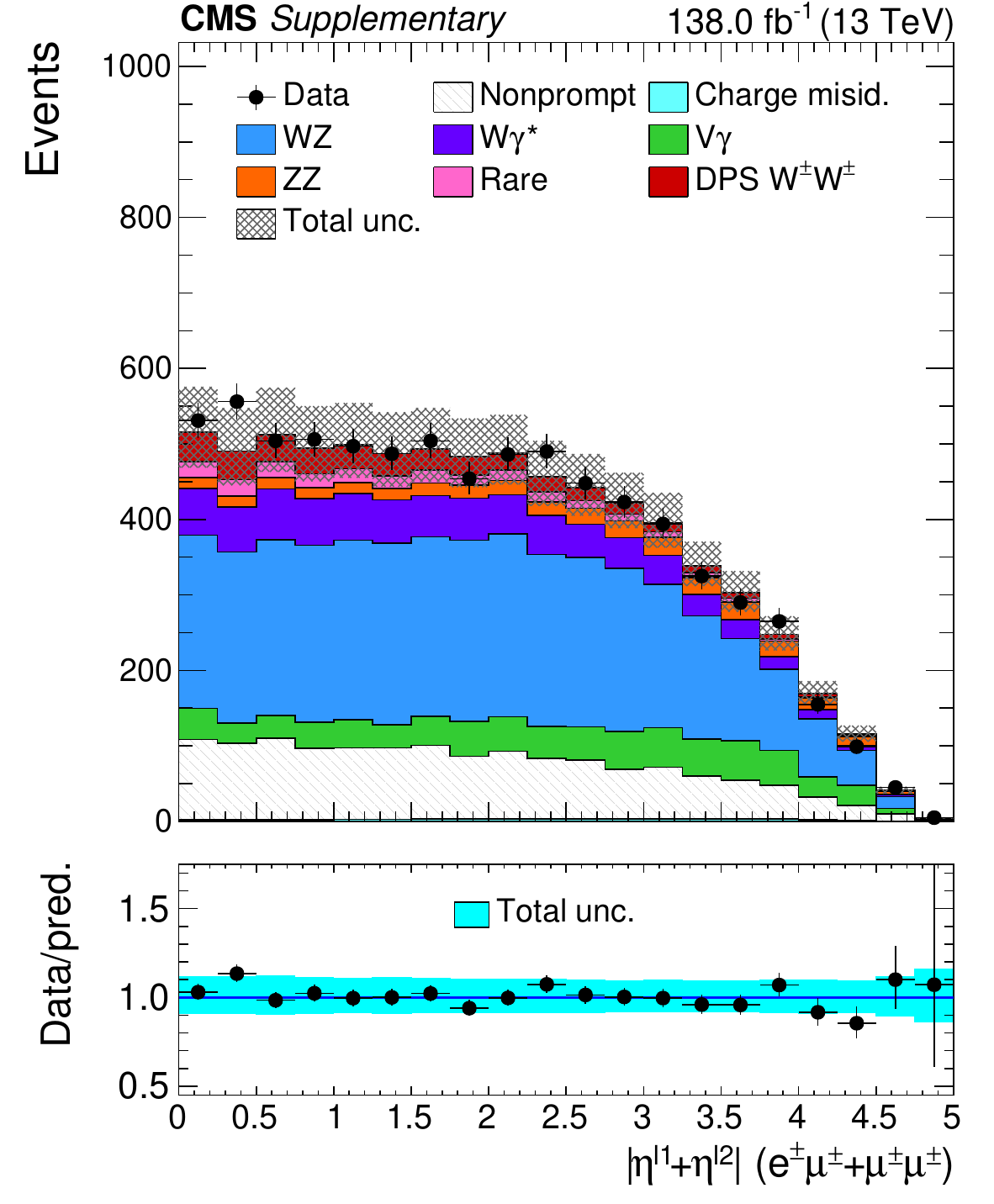}\\%etaS
\caption{Distributions of the kinematic variables used for the training of the BDT discriminants for the combined \emu and \mumu final states: \ptone, \pttwo, \ptmiss, \mttwo, \mtll, \mtone, \dphill, \dphiltwomet, \dphillltwo, \etaprod, and, \etasum. The signal and background yields have been normalized to their respective postfit yields. The uncertainty bands represent the total expected uncertainty on the predicted yields, which includes both the statistical and systematic components.}\label{fig:btdIVprost}
\end{figure*}
\begin{figure*}[htbp!]
\centering
\includegraphics[width=0.44\textwidth]{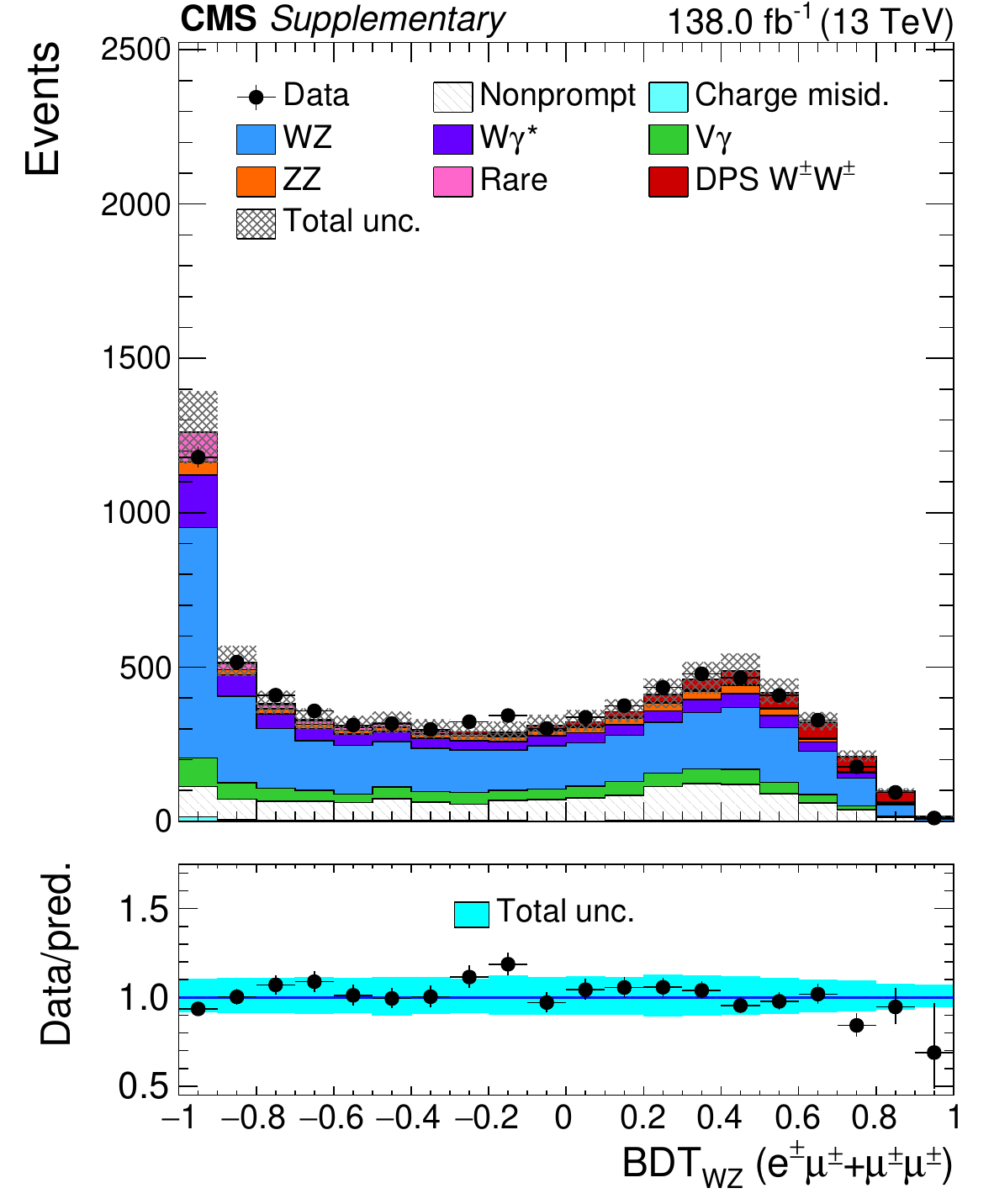}%BDTwz  
\includegraphics[width=0.44\textwidth]{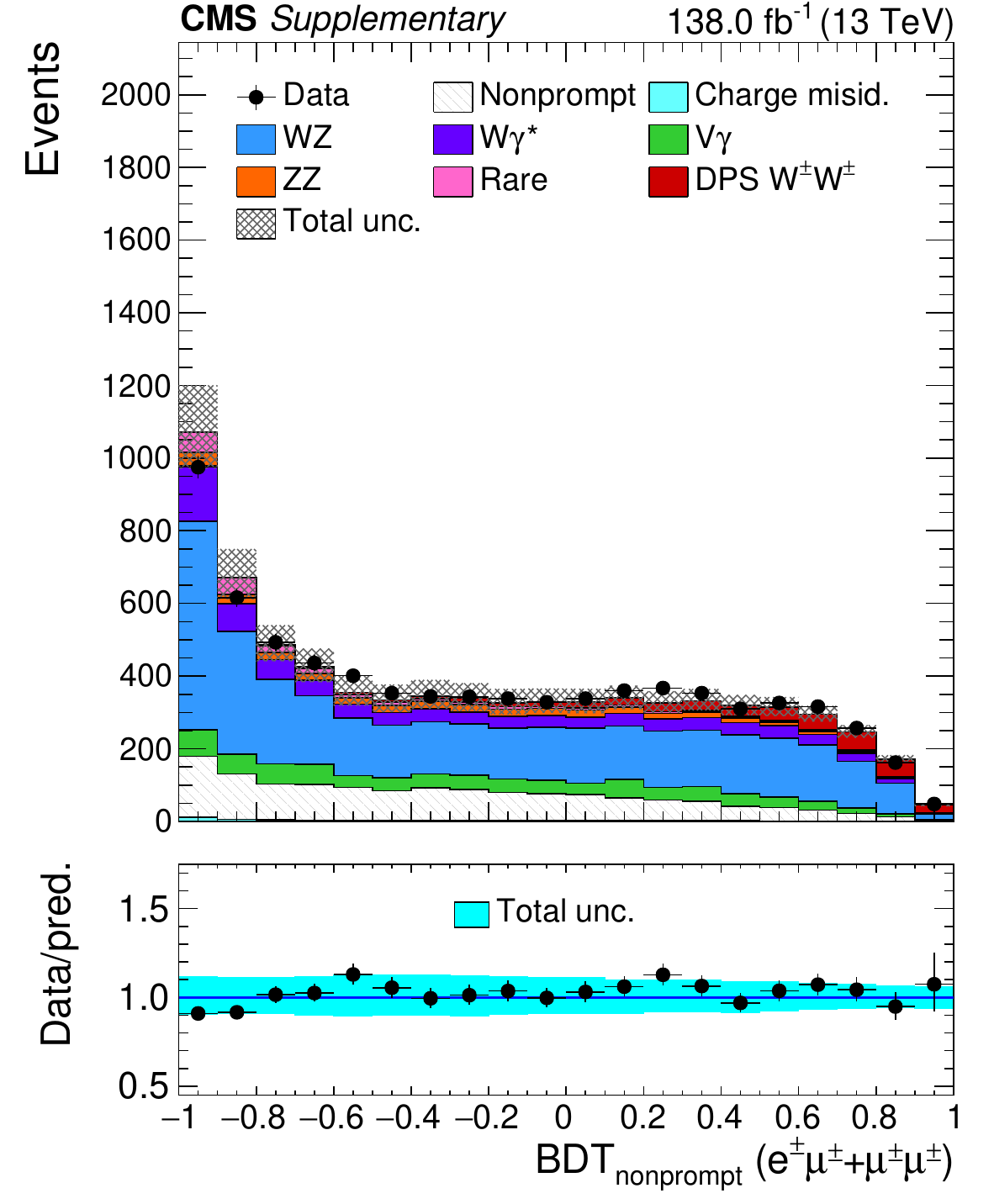}%BDTnp
\caption{Distributions of the two single BDT discriminants for the combined \emu and \mumu final states. The BDT discriminant trained against the \WZ (nonprompt leptons) background is labeled as BDT$_{\WZ}$ (BDT$_{\rm nonprompt}$). The signal and background yields have been normalized to their respective postfit yields. The uncertainty bands represent the total expected uncertainty on the predicted yields, which includes both the statistical and systematic components.}\label{fig:btdprost}
\end{figure*}

%% file: SMP-21-013-public-authorlist.tex
\cmsinstitute{Yerevan Physics Institute, Yerevan, Armenia}
{\tolerance=6000
A.~Tumasyan\cmsAuthorMark{1}\cmsorcid{0009-0000-0684-6742}
\par}
\cmsinstitute{Institut f\"{u}r Hochenergiephysik, Vienna, Austria}
{\tolerance=6000
W.~Adam\cmsorcid{0000-0001-9099-4341}, J.W.~Andrejkovic, T.~Bergauer\cmsorcid{0000-0002-5786-0293}, S.~Chatterjee\cmsorcid{0000-0003-2660-0349}, K.~Damanakis\cmsorcid{0000-0001-5389-2872}, M.~Dragicevic\cmsorcid{0000-0003-1967-6783}, A.~Escalante~Del~Valle\cmsorcid{0000-0002-9702-6359}, P.S.~Hussain\cmsorcid{0000-0002-4825-5278}, M.~Jeitler\cmsAuthorMark{2}\cmsorcid{0000-0002-5141-9560}, N.~Krammer\cmsorcid{0000-0002-0548-0985}, L.~Lechner\cmsorcid{0000-0002-3065-1141}, D.~Liko\cmsorcid{0000-0002-3380-473X}, I.~Mikulec\cmsorcid{0000-0003-0385-2746}, P.~Paulitsch, F.M.~Pitters, J.~Schieck\cmsAuthorMark{2}\cmsorcid{0000-0002-1058-8093}, R.~Sch\"{o}fbeck\cmsorcid{0000-0002-2332-8784}, D.~Schwarz\cmsorcid{0000-0002-3821-7331}, S.~Templ\cmsorcid{0000-0003-3137-5692}, W.~Waltenberger\cmsorcid{0000-0002-6215-7228}, C.-E.~Wulz\cmsAuthorMark{2}\cmsorcid{0000-0001-9226-5812}
% ...
\par}
\vskip\cmsinstskip
\dag:~Deceased\\
$^{1}$Also at Yerevan State University, Yerevan, Armenia\\
$^{2}$Also at TU Wien, Vienna, Austria\\
% ...


%% file: SMP-21-013_temp.bbl
\providecommand{\href}[2]{#2}\begingroup\raggedright\begin{thebibliography}{10}%
\makeatletter
\providecommand{\hrefCMSnoop }[0]{\@secondoftwo}%
\makeatother
\providecommand{\doi}{\texttt{doi:}\begingroup \urlstyle{tt}\Url}

\bibitem{Paver:1982yp}
\hrefCMSnoop {}{N.~Paver and D.~Treleani, ``{Multiquark scattering and
  large-$p_{\rm T}$ jet production in hadronic collisions}'',} \textit{ Nuovo
  Cim. A} \textbf{ 70} (1982) 215,
  \href{http://dx.doi.org/10.1007/BF02814035}{\doi{10.1007/BF02814035}}.

\bibitem{Goebel:1979mi}
\hrefCMSnoop {}{C.~Goebel, D.~M. Scott, and F.~Halzen, ``{Double {Drell-Yan}
  annihilations in hadron collisions: novel tests of the constituent
  picture}'',} \textit{ Phys. Rev. D} \textbf{ 22} (1980) 2789,
\href{http://dx.doi.org/10.1103/PhysRevD.22.2789}{\doi{10.1103/PhysRevD.22.2789}}.
%%CITATION = PHRVA,D22,2789;%%.

\bibitem{Shelest:1982dg}
\hrefCMSnoop {}{V.~P. Shelest, A.~M. Snigirev, and G.~M. Zinovev, ``{Gazing
  into the multiparton distribution equations in {QCD}}'',} \textit{ Phys.
  Lett. B} \textbf{ 113} (1982) 325,
\href{http://dx.doi.org/10.1016/0370-2693(82)90049-1}{\doi{10.1016/0370-2693(82)90049-1}}.
%%CITATION = PHLTA,113B,325;%%.

\bibitem{Sjostrand:1987su}
\hrefCMSnoop {}{T.~Sj{\"o}strand and M.~van Zijl, ``{A multiple interaction
  model for the event structure in hadron collisions}'',} \textit{ Phys. Rev.
  D} \textbf{ 36} (1987) 2019,
\href{http://dx.doi.org/10.1103/PhysRevD.36.2019}{\doi{10.1103/PhysRevD.36.2019}}.
%%CITATION = PHRVA,D36,2019;%%.

\bibitem{dps2}
\hrefCMSnoop {}{G.~Calucci and D.~Treleani, ``{Disentangling correlations in
  multiple parton interactions}'',} \textit{ Phys. Rev. D} \textbf{ 83} (2011)
  016012,
  \href{http://dx.doi.org/10.1103/PhysRevD.83.016012}{\doi{10.1103/PhysRevD.83.016012}},
\href{http://www.arXiv.org/abs/1009.5881}{\texttt{arXiv:1009.5881}}.
%%CITATION = ARXIV:1009.5881;%%.

\bibitem{dps1}
\hrefCMSnoop {}{M.~Diehl, D.~Ostermeier, and A.~Schafer, ``{Elements of a
  theory for multiparton interactions in QCD}'',} \textit{ JHEP} \textbf{ 03}
  (2012) 089,
  \href{http://dx.doi.org/10.1007/JHEP03(2012)089}{\doi{10.1007/JHEP03(2012)089}},
\href{http://www.arXiv.org/abs/1111.0910}{\texttt{arXiv:1111.0910}}.
%%CITATION = ARXIV:1111.0910;%%.

\bibitem{Blok:2013bpa}
\hrefCMSnoop {}{B.~Blok, {\relax Yu}.~Dokshitzer, L.~Frankfurt, and
  M.~Strikman, ``{Perturbative QCD correlations in multi-parton collisions}'',}
  \textit{ Eur. Phys. J. C} \textbf{ 74} (2014) 2926,
  \href{http://dx.doi.org/10.1140/epjc/s10052-014-2926-z}{\doi{10.1140/epjc/s10052-014-2926-z}},
\href{http://www.arXiv.org/abs/1306.3763}{\texttt{arXiv:1306.3763}}.
%%CITATION = ARXIV:1306.3763;%%.

\bibitem{dpsTheory}
\hrefCMSnoop {}{M.~Diehl and J.~R. Gaunt, ``{Double parton scattering theory
  overview}'',} \textit{ Adv. Ser. Direct. High Energy Phys.} \textbf{ 29}
  (2018) 7,
  \href{http://dx.doi.org/10.1142/9789813227767_0002}{\doi{10.1142/9789813227767_0002}},
\href{http://www.arXiv.org/abs/1710.04408}{\texttt{arXiv:1710.04408}}.
%%CITATION = ARXIV:1710.04408;%%.

\bibitem{Bartalini:2017jkk}
P.~Bartalini and J.~R. Gaunt, eds., ``{Multiple parton interactions at the
  LHC}'', volume~29 of \textit{ Advanced series on directions in high energy
  physics}.
\newblock World Scientific, 2018.
\newblock \href{http://dx.doi.org/10.1142/10646}{\doi{10.1142/10646}},
  ISBN~978-981-322-775-0, 978-981-322-777-4.

\bibitem{Gaunt}
\hrefCMSnoop {}{J.~R. Gaunt, C.-H. Kom, A.~Kulesza, and W.~J. Stirling,
  ``{Same-sign W pair production as a probe of double parton scattering at the
  LHC}'',} \textit{ Eur. Phys. J. C} \textbf{ 69} (2010) 53,
  \href{http://dx.doi.org/10.1140/epjc/s10052-010-1362-y}{\doi{10.1140/epjc/s10052-010-1362-y}},
\href{http://www.arXiv.org/abs/1003.3953}{\texttt{arXiv:1003.3953}}.
%%CITATION = ARXIV:1003.3953;%%.

\bibitem{sigeff2}
\hrefCMSnoop {}{{ATLAS Collaboration}, ``{Measurement of hard double-parton
  interactions in W$(\to {l\nu})$+ 2 jet events at $\sqrt{s}$ = 7 TeV with the
  ATLAS detector}'',} \textit{ New J. Phys.} \textbf{ 15} (2013) 033038,
  \href{http://dx.doi.org/10.1088/1367-2630/15/3/033038}{\doi{10.1088/1367-2630/15/3/033038}},
\href{http://www.arXiv.org/abs/1301.6872}{\texttt{arXiv:1301.6872}}.
%%CITATION = ARXIV:1301.6872;%%.

\bibitem{sigeff1}
\hrefCMSnoop {}{{CMS Collaboration}, ``{Study of double parton scattering using
  W + 2-jet events in proton-proton collisions at $\sqrt{s}$ = 7 TeV}'',}
  \textit{ JHEP} \textbf{ 03} (2014) 032,
  \href{http://dx.doi.org/10.1007/JHEP03(2014)032}{\doi{10.1007/JHEP03(2014)032}},
\href{http://www.arXiv.org/abs/1312.5729}{\texttt{arXiv:1312.5729}}.
%%CITATION = ARXIV:1312.5729;%%.

\bibitem{sigeff4}
\hrefCMSnoop {}{{CDF} Collaboration, ``{Measurement of double parton scattering
  in $\overline{\mathrm{p}}$p collisions at $\sqrt{s}$ = 1.8 TeV}'',} \textit{
  Phys. Rev. Lett.} \textbf{ 79} (1997) 584,
  \href{http://dx.doi.org/10.1103/PhysRevLett.79.584}{\doi{10.1103/PhysRevLett.79.584}}.

\bibitem{sigeff7}
\hrefCMSnoop {}{{CMS Collaboration}, ``{Event generator tunes obtained from
  underlying event and multiparton scattering measurements}'',} \textit{ Eur.
  Phys. J. C} \textbf{ 76} (2016) 155,
  \href{http://dx.doi.org/10.1140/epjc/s10052-016-3988-x}{\doi{10.1140/epjc/s10052-016-3988-x}},
\href{http://www.arXiv.org/abs/1512.00815}{\texttt{arXiv:1512.00815}}.
%%CITATION = ARXIV:1512.00815;%%.

\bibitem{sigeff3}
\hrefCMSnoop {}{{D0} Collaboration, ``{Double parton interactions in
  $\ensuremath{\gamma}+3$ jet events in p$\overline{\rm p}$ collisions at
  $\sqrt{s}=1.96$ TeV}'',} \textit{ Phys. Rev. D} \textbf{ 81} (2010) 052012,
  \href{http://dx.doi.org/10.1103/PhysRevD.81.052012}{\doi{10.1103/PhysRevD.81.052012}},
  \href{http://www.arXiv.org/abs/0912.5104}{\texttt{arXiv:0912.5104}}.

\bibitem{Aad:2014kba}
\hrefCMSnoop {}{{ATLAS Collaboration}, ``{Observation and measurements of the
  production of prompt and non-prompt $\JPsi$ mesons in association with a Z
  boson in pp collisions at $\sqrt{s}$ = 8 TeV with the ATLAS detector}'',}
  \textit{ Eur. Phys. J. C} \textbf{ 75} (2015) 229,
  \href{http://dx.doi.org/10.1140/epjc/s10052-015-3406-9}{\doi{10.1140/epjc/s10052-015-3406-9}},
\href{http://www.arXiv.org/abs/1412.6428}{\texttt{arXiv:1412.6428}}.
%%CITATION = ARXIV:1412.6428;%%.

\bibitem{Aaij:2015wpa}
\hrefCMSnoop {}{{LHCb Collaboration}, ``{Production of associated $\Upsilon$
  and open charm hadrons in pp collisions at $\sqrt{s}=7 $ and 8 TeV via double
  parton scattering}'',} \textit{ JHEP} \textbf{ 07} (2016) 052,
  \href{http://dx.doi.org/10.1007/JHEP07(2016)052}{\doi{10.1007/JHEP07(2016)052}},
\href{http://www.arXiv.org/abs/1510.05949}{\texttt{arXiv:1510.05949}}.
%%CITATION = ARXIV:1510.05949;%%.

\bibitem{Abazov:2015fbl}
\hrefCMSnoop {}{{D0} Collaboration, ``{Evidence for simultaneous production of
  $\JPsi$ and $\Upsilon$ mesons}'',} \textit{ Phys. Rev. Lett.} \textbf{ 116}
  (2016) 082002,
  \href{http://dx.doi.org/10.1103/PhysRevLett.116.082002}{\doi{10.1103/PhysRevLett.116.082002}},
\href{http://www.arXiv.org/abs/1511.02428}{\texttt{arXiv:1511.02428}}.
%%CITATION = ARXIV:1511.02428;%%.

\bibitem{CMS:2021qsn}
\hrefCMSnoop {}{{CMS Collaboration}, ``{Observation of triple J/$\psi$ meson
  production in proton-proton collisions at $\sqrt{s}$ = 13 TeV}'',} 2021.
  \href{http://www.arXiv.org/abs/2111.05370}{\texttt{arXiv:2111.05370}}.
  Submitted to \textit{Nature Physics}.

\bibitem{sigeff5}
\hrefCMSnoop {}{{AFS} Collaboration, ``{Double parton scattering in pp
  collisions at $\sqrt{s}=63$ {GeV}}'',} \textit{ Z. Phys. C} \textbf{ 34}
  (1987) 163,
\href{http://dx.doi.org/10.1007/BF01566757}{\doi{10.1007/BF01566757}}.
%%CITATION = ZEPYA,C34,163;%%.

\bibitem{sigeff6}
\hrefCMSnoop {}{{UA2} Collaboration, ``{A study of multi-jet events at the CERN
  $\overline{\rm p}$p collider and a search for double parton scattering}'',}
  \textit{ Phys. Lett. B} \textbf{ 268} (1991) 145,
\href{http://dx.doi.org/10.1016/0370-2693(91)90937-L}{\doi{10.1016/0370-2693(91)90937-L}}.
%%CITATION = PHLTA,B268,145;%%.

\bibitem{cdf:dps}
\hrefCMSnoop {}{{CDF} Collaboration, ``{Study of four-jet events and evidence
  for double parton interactions in p$\overline{\rm p}$ collisions at
  $\sqrt{s}$ = 1.8 TeV}'',} \textit{ Phys. Rev. D} \textbf{ 47} (1993) 4857,
  \href{http://dx.doi.org/10.1103/PhysRevD.47.4857}{\doi{10.1103/PhysRevD.47.4857}}.

\bibitem{CDF:dps2}
\hrefCMSnoop {}{{CDF} Collaboration, ``{Double parton scattering in
  $\overline{\mathrm{p}}$p collisions at $\sqrt{s}$ = 1.8 TeV}'',} \textit{
  Phys. Rev. D} \textbf{ 56} (1997) 3811,
  \href{http://dx.doi.org/10.1103/PhysRevD.56.3811}{\doi{10.1103/PhysRevD.56.3811}}.

\bibitem{LHCb:dps}
\hrefCMSnoop {}{{LHCb Collaboration}, ``{Observation of double charm production
  involving open charm in pp collisions at $\sqrt{s}$ = 7 TeV}'',} \textit{
  JHEP} \textbf{ 06} (2012) 141,
  \href{http://dx.doi.org/10.1007/JHEP06(2012)141}{\doi{10.1007/JHEP06(2012)141}},
  \href{http://www.arXiv.org/abs/1205.0975}{\texttt{arXiv:1205.0975}}.
[Addendum: \DOI{10.1007/JHEP03(2014)108}].
%%CITATION = ARXIV:1205.0975;%%.

\bibitem{CMS-FSQ-16-005}
\hrefCMSnoop {}{{CMS Collaboration}, ``{Constraints on the double-parton
  scattering cross section from same-sign W boson pair production in
  proton-proton collisions at $ \sqrt{s}=8 $ TeV}'',} \textit{ JHEP} \textbf{
  02} (2018) 032,
  \href{http://dx.doi.org/10.1007/JHEP02(2018)032}{\doi{10.1007/JHEP02(2018)032}},
\href{http://www.arXiv.org/abs/1712.02280}{\texttt{arXiv:1712.02280}}.
%%CITATION = ARXIV:1712.02280;%%.

\bibitem{atlas:dps2018}
\hrefCMSnoop {}{{ATLAS Collaboration}, ``{Study of the hard double-parton
  scattering contribution to inclusive four-lepton production in pp collisions
  at $\sqrt s=$ 8 TeV with the ATLAS detector}'',} \textit{ Phys. Lett. B}
  \textbf{ 790} (2019) 595,
  \href{http://dx.doi.org/10.1016/j.physletb.2019.01.062}{\doi{10.1016/j.physletb.2019.01.062}},
\href{http://www.arXiv.org/abs/1811.11094}{\texttt{arXiv:1811.11094}}.
%%CITATION = ARXIV:1811.11094;%%.

\bibitem{CMS:2021lxi}
\hrefCMSnoop {}{{CMS Collaboration}, ``Measurement of double-parton scattering
  in inclusive production of four jets with low transverse momentum in
  proton-proton collisions at $\sqrt{s} = 13$ {TeV}'',} \textit{ JHEP} \textbf{
  01} (2022) 177,
  \href{http://dx.doi.org/10.1007/JHEP01(2022)177}{\doi{10.1007/JHEP01(2022)177}},
  \href{http://www.arXiv.org/abs/2109.13822}{\texttt{arXiv:2109.13822}}.

\bibitem{CMS:2021wfx}
\hrefCMSnoop {}{{CMS Collaboration}, ``{Study of Z boson plus jets events using
  variables sensitive to double-parton scattering in pp collisions at 13
  TeV}'',} \textit{ JHEP} \textbf{ 10} (2021) 176,
  \href{http://dx.doi.org/10.1007/JHEP10(2021)176}{\doi{10.1007/JHEP10(2021)176}},
  \href{http://www.arXiv.org/abs/2105.14511}{\texttt{arXiv:2105.14511}}.

\bibitem{D0:2014vql}
\hrefCMSnoop {}{{D0} Collaboration, ``{Observation and studies of double
  $\JPsi$ production at the Tevatron}'',} \textit{ Phys. Rev. D} \textbf{ 90}
  (2014) 111101,
  \href{http://dx.doi.org/10.1103/PhysRevD.90.111101}{\doi{10.1103/PhysRevD.90.111101}},
  \href{http://www.arXiv.org/abs/1406.2380}{\texttt{arXiv:1406.2380}}.

\bibitem{CMS:2020cpy}
\hrefCMSnoop {}{{CMS Collaboration}, ``{Search for physics beyond the standard
  model in events with jets and two same-sign or at least three charged leptons
  in proton-proton collisions at $\sqrt{s}=$ 13 TeV}'',} \textit{ Eur. Phys. J.
  C} \textbf{ 80} (2020) 752,
  \href{http://dx.doi.org/10.1140/epjc/s10052-020-8168-3}{\doi{10.1140/epjc/s10052-020-8168-3}},
  \href{http://www.arXiv.org/abs/2001.10086}{\texttt{arXiv:2001.10086}}.

\bibitem{DelFabbro:1999tf}
\hrefCMSnoop {}{A.~Del~Fabbro and D.~Treleani, ``{Double parton scattering
  background to Higgs boson production at the LHC}'',} \textit{ Phys. Rev. D}
  \textbf{ 61} (2000) 077502,
  \href{http://dx.doi.org/10.1103/PhysRevD.61.077502}{\doi{10.1103/PhysRevD.61.077502}},
  \href{http://www.arXiv.org/abs/hep-ph/9911358}{\texttt{arXiv:hep-ph/9911358}}.

\bibitem{Gaunt:2009re}
\hrefCMSnoop {}{J.~R. Gaunt and W.~J. Stirling, ``{Double parton distributions
  incorporating perturbative QCD evolution and momentum and quark number sum
  rules}'',} \textit{ JHEP} \textbf{ 03} (2010) 005,
  \href{http://dx.doi.org/10.1007/JHEP03(2010)005}{\doi{10.1007/JHEP03(2010)005}},
  \href{http://www.arXiv.org/abs/0910.4347}{\texttt{arXiv:0910.4347}}.

\bibitem{Diehl:2017kgu}
\hrefCMSnoop {}{M.~Diehl, J.~R. Gaunt, and K.~Sch{\"o}nwald, ``{Double hard
  scattering without double counting}'',} \textit{ JHEP} \textbf{ 06} (2017)
  083,
  \href{http://dx.doi.org/10.1007/JHEP06(2017)083}{\doi{10.1007/JHEP06(2017)083}},
  \href{http://www.arXiv.org/abs/1702.06486}{\texttt{arXiv:1702.06486}}.

\bibitem{Diehl:2018kgr}
\hrefCMSnoop {}{M.~Diehl, P.~Pl{\"o}{\ss}{}l, and A.~Sch{\"a}fer, ``{Proof of
  sum rules for double parton distributions in QCD}'',} \textit{ Eur. Phys. J.
  C} \textbf{ 79} (2019) 253,
  \href{http://dx.doi.org/10.1140/epjc/s10052-019-6777-5}{\doi{10.1140/epjc/s10052-019-6777-5}},
  \href{http://www.arXiv.org/abs/1811.00289}{\texttt{arXiv:1811.00289}}.

\bibitem{Gaunt:2018eix}
\hrefCMSnoop {}{J.~R. Gaunt and T.~Kasemets, ``{Transverse momentum dependence
  in double parton scattering}'',} \textit{ Adv. High Energy Phys.} \textbf{
  2019} (2019) 3797394,
  \href{http://dx.doi.org/10.1155/2019/3797394}{\doi{10.1155/2019/3797394}},
  \href{http://www.arXiv.org/abs/1812.09099}{\texttt{arXiv:1812.09099}}.

\bibitem{Cabouat:2019gtm}
\hrefCMSnoop {}{B.~Cabouat, J.~R. Gaunt, and K.~Ostrolenk, ``{A Monte Carlo
  simulation of double parton scattering}'',} \textit{ JHEP} \textbf{ 11}
  (2019) 061,
  \href{http://dx.doi.org/10.1007/JHEP11(2019)061}{\doi{10.1007/JHEP11(2019)061}},
  \href{http://www.arXiv.org/abs/1906.04669}{\texttt{arXiv:1906.04669}}.

\bibitem{Kulesza:1999zh}
\hrefCMSnoop {}{A.~Kulesza and W.~J. Stirling, ``{Like sign W boson production
  at the LHC as a probe of double parton scattering}'',} \textit{ Phys. Lett.
  B} \textbf{ 475} (2000) 168,
  \href{http://dx.doi.org/10.1016/S0370-2693(99)01512-9}{\doi{10.1016/S0370-2693(99)01512-9}},
\href{http://www.arXiv.org/abs/hep-ph/9912232}{\texttt{arXiv:hep-ph/9912232}}.
%%CITATION = HEP-PH/9912232;%%.

\bibitem{Sirunyan:2019zox}
\hrefCMSnoop {}{{CMS Collaboration}, ``{Evidence for WW production from
  double-parton interactions in proton-proton collisions at $\sqrt{s} = 13$
  TeV}'',} \textit{ Eur. Phys. J. C} \textbf{ 80} (2020) 41,
  \href{http://dx.doi.org/10.1140/epjc/s10052-019-7541-6}{\doi{10.1140/epjc/s10052-019-7541-6}},
  \href{http://www.arXiv.org/abs/1909.06265}{\texttt{arXiv:1909.06265}}.

\bibitem{Chatrchyan:2008zzk}
\hrefCMSnoop {}{{CMS Collaboration}, ``The {CMS} experiment at the {CERN}
  {LHC}'',} \textit{ JINST} \textbf{ 3} (2008) S08004,
\href{http://dx.doi.org/10.1088/1748-0221/3/08/S08004}{\doi{10.1088/1748-0221/3/08/S08004}}.
%%CITATION = JINST,3,S08004;%%.

\bibitem{CMS:2020cmk}
\hrefCMSnoop {}{{CMS Collaboration}, ``{Performance of the CMS Level-1 trigger
  in proton-proton collisions at $\sqrt{s} = 13$\,TeV}'',} \textit{ JINST}
  \textbf{ 15} (2020) P10017,
  \href{http://dx.doi.org/10.1088/1748-0221/15/10/P10017}{\doi{10.1088/1748-0221/15/10/P10017}},
  \href{http://www.arXiv.org/abs/2006.10165}{\texttt{arXiv:2006.10165}}.

\bibitem{Khachatryan:2016bia}
\hrefCMSnoop {}{{CMS Collaboration}, ``{The CMS trigger system}'',} \textit{
  JINST} \textbf{ 12} (2017) P01020,
  \href{http://dx.doi.org/10.1088/1748-0221/12/01/P01020}{\doi{10.1088/1748-0221/12/01/P01020}},
\href{http://www.arXiv.org/abs/1609.02366}{\texttt{arXiv:1609.02366}}.
%%CITATION = ARXIV:1609.02366;%%.

\bibitem{Khachatryan:2015hwa}
\hrefCMSnoop {}{{CMS Collaboration}, ``{Performance of electron reconstruction
  and selection with the CMS detector in proton-proton collisions at $\sqrt{s}
  = 8$\TeV}'',} \textit{ JINST} \textbf{ 10} (2015) P06005,
  \href{http://dx.doi.org/10.1088/1748-0221/10/06/P06005}{\doi{10.1088/1748-0221/10/06/P06005}},
\href{http://www.arXiv.org/abs/1502.02701}{\texttt{arXiv:1502.02701}}.
%%CITATION = ARXIV:1502.02701;%%.

\bibitem{Sirunyan:2018fpa}
\hrefCMSnoop {}{{CMS Collaboration}, ``{Performance of the CMS muon detector
  and muon reconstruction with proton-proton collisions at $\sqrt{s}=$ 13
  TeV}'',} \textit{ JINST} \textbf{ 13} (2018) P06015,
  \href{http://dx.doi.org/10.1088/1748-0221/13/06/P06015}{\doi{10.1088/1748-0221/13/06/P06015}},
\href{http://www.arXiv.org/abs/1804.04528}{\texttt{arXiv:1804.04528}}.
%%CITATION = ARXIV:1804.04528;%%.

\bibitem{CMS:2015myp}
\hrefCMSnoop {}{{CMS Collaboration}, ``Performance of photon reconstruction and
  identification with the {CMS} detector in proton-proton collisions at
  $\sqrt{s} = 8$ {TeV}'',} \textit{ JINST} \textbf{ 10} (2015) P08010,
  \href{http://dx.doi.org/10.1088/1748-0221/10/08/P08010}{\doi{10.1088/1748-0221/10/08/P08010}},
  \href{http://www.arXiv.org/abs/1502.02702}{\texttt{arXiv:1502.02702}}.

\bibitem{CMS:2014pgm}
\hrefCMSnoop {}{{CMS Collaboration}, ``{Description and performance of track
  and primary-vertex reconstruction with the CMS tracker}'',} \textit{ JINST}
  \textbf{ 9} (2014) P10009,
  \href{http://dx.doi.org/10.1088/1748-0221/9/10/P10009}{\doi{10.1088/1748-0221/9/10/P10009}},
  \href{http://www.arXiv.org/abs/1405.6569}{\texttt{arXiv:1405.6569}}.

\bibitem{Sirunyan:2017ulk}
\hrefCMSnoop {}{{CMS Collaboration}, ``{Particle-flow reconstruction and global
  event description with the CMS detector}'',} \textit{ JINST} \textbf{ 12}
  (2017) P10003,
  \href{http://dx.doi.org/10.1088/1748-0221/12/10/P10003}{\doi{10.1088/1748-0221/12/10/P10003}},
\href{http://www.arXiv.org/abs/1706.04965}{\texttt{arXiv:1706.04965}}.
%%CITATION = ARXIV:1706.04965;%%.

\bibitem{Sirunyan:2018pgf}
\hrefCMSnoop {}{{CMS Collaboration}, ``{Performance of reconstruction and
  identification of $\tau$ leptons decaying to hadrons and $\nu_{\tau}$ in pp
  collisions at $\sqrt{s}=$ 13 TeV}'',} \textit{ JINST} \textbf{ 13} (2018)
  P10005,
  \href{http://dx.doi.org/10.1088/1748-0221/13/10/P10005}{\doi{10.1088/1748-0221/13/10/P10005}},
\href{http://www.arXiv.org/abs/1809.02816}{\texttt{arXiv:1809.02816}}.
%%CITATION = ARXIV:1809.02816;%%.

\bibitem{Khachatryan:2016kdb}
\hrefCMSnoop {}{{CMS Collaboration}, ``{Jet energy scale and resolution in the
  CMS experiment in pp collisions at 8 TeV}'',} \textit{ JINST} \textbf{ 12}
  (2017) P02014,
  \href{http://dx.doi.org/10.1088/1748-0221/12/02/P02014}{\doi{10.1088/1748-0221/12/02/P02014}},
\href{http://www.arXiv.org/abs/1607.03663}{\texttt{arXiv:1607.03663}}.
%%CITATION = ARXIV:1607.03663;%%.

\bibitem{CMS:2019ctu}
\hrefCMSnoop {}{{CMS Collaboration}, ``Performance of missing transverse
  momentum reconstruction in proton-proton collisions at $\sqrt{s} = 13$\,{TeV}
  using the {CMS} detector'',} \textit{ JINST} \textbf{ 14} (2019) P07004,
  \href{http://dx.doi.org/10.1088/1748-0221/14/07/P07004}{\doi{10.1088/1748-0221/14/07/P07004}},
\href{http://www.arXiv.org/abs/1903.06078}{\texttt{arXiv:1903.06078}}.
%%CITATION = ARXIV:1903.06078;%%.

\bibitem{Cacciari:2008gp}
\hrefCMSnoop {}{M.~Cacciari, G.~P. Salam, and G.~Soyez, ``{The anti-\kt jet
  clustering algorithm}'',} \textit{ JHEP} \textbf{ 04} (2008) 063,
  \href{http://dx.doi.org/10.1088/1126-6708/2008/04/063}{\doi{10.1088/1126-6708/2008/04/063}},
  \href{http://www.arXiv.org/abs/0802.1189}{\texttt{arXiv:0802.1189}}.

\bibitem{FastJet}
\hrefCMSnoop {}{M.~Cacciari, G.~P. Salam, and G.~Soyez, ``{FastJet} user
  manual'',} \textit{ Eur. Phys. J. C} \textbf{ 72} (2012) 1896,
  \href{http://dx.doi.org/10.1140/epjc/s10052-012-1896-2}{\doi{10.1140/epjc/s10052-012-1896-2}},
\href{http://www.arXiv.org/abs/1111.6097}{\texttt{arXiv:1111.6097}}.
%%CITATION = ARXIV:1111.6097;%%.

\bibitem{CMS:2020ebo}
\hrefCMSnoop {}{{CMS Collaboration}, ``{Pileup mitigation at CMS in 13 TeV
  data}'',} \textit{ JINST} \textbf{ 15} (2020) P09018,
  \href{http://dx.doi.org/10.1088/1748-0221/15/09/P09018}{\doi{10.1088/1748-0221/15/09/P09018}},
  \href{http://www.arXiv.org/abs/2003.00503}{\texttt{arXiv:2003.00503}}.

\bibitem{Sirunyan:2020ycc}
\hrefCMSnoop {}{{CMS Collaboration}, ``{Electron and photon reconstruction and
  identification with the CMS experiment at the CERN LHC}'',} \textit{ JINST}
  \textbf{ 16} (2021) P05014,
  \href{http://dx.doi.org/10.1088/1748-0221/16/05/P05014}{\doi{10.1088/1748-0221/16/05/P05014}},
  \href{http://www.arXiv.org/abs/2012.06888}{\texttt{arXiv:2012.06888}}.

\bibitem{CMS:2020mpn}
\hrefCMSnoop {}{{CMS Collaboration}, ``{Measurement of the Higgs boson
  production rate in association with top quarks in final states with
  electrons, muons, and hadronically decaying tau leptons at $\sqrt{s} =$ 13
  TeV}'',} \textit{ Eur. Phys. J. C} \textbf{ 81} (2021) 378,
  \href{http://dx.doi.org/10.1140/epjc/s10052-021-09014-x}{\doi{10.1140/epjc/s10052-021-09014-x}},
  \href{http://www.arXiv.org/abs/2011.03652}{\texttt{arXiv:2011.03652}}.

\bibitem{Bols:2020bkb}
E.~Bols\hrefCMSnoop {}{ {et~al.}, ``Jet flavour classification using
  {DeepJet}'',} \textit{ JINST} \textbf{ 15} (2020) P12012,
  \href{http://dx.doi.org/10.1088/1748-0221/15/12/P12012}{\doi{10.1088/1748-0221/15/12/P12012}},
  \href{http://www.arXiv.org/abs/2008.10519}{\texttt{arXiv:2008.10519}}.

\bibitem{CMS-DP-2018-058}
\href {http://cds.cern.ch/record/2646773}{{CMS Collaboration}, ``{Performance
  of the DeepJet b tagging algorithm using 41.9 fb$^{-1}$ of data from
  proton-proton collisions at 13 TeV with Phase 1 CMS detector}'',} CMS
  Detector Performance Note CMS-DP-2018-058, CERN, 2018.

\bibitem{Sjostrand:2014zea}
T.~Sj{\"o}strand\hrefCMSnoop {}{ {et~al.}, ``{An introduction to PYTHIA
  8.2}'',} \textit{ Comput. Phys. Commun.} \textbf{ 191} (2015) 159,
  \href{http://dx.doi.org/10.1016/j.cpc.2015.01.024}{\doi{10.1016/j.cpc.2015.01.024}},
\href{http://www.arXiv.org/abs/1410.3012}{\texttt{arXiv:1410.3012}}.
%%CITATION = ARXIV:1410.3012;%%.

\bibitem{Bahr:2008pv}
\hrefCMSnoop {}{M.~Bahr {et~al.}, ``{Herwig++ physics and manual}'',} \textit{
  Eur. Phys. J. C} \textbf{ 58} (2008) 639,
  \href{http://dx.doi.org/10.1140/epjc/s10052-008-0798-9}{\doi{10.1140/epjc/s10052-008-0798-9}},
\href{http://www.arXiv.org/abs/0803.0883}{\texttt{arXiv:0803.0883}}.
%%CITATION = ARXIV:0803.0883;%%.

\bibitem{Ball:2012cx}
\hrefCMSnoop {}{R.~D. Ball {et~al.}, ``{Parton distributions with LHC data}'',}
  \textit{ Nucl. Phys. B} \textbf{ 867} (2013) 244,
  \href{http://dx.doi.org/10.1016/j.nuclphysb.2012.10.003}{\doi{10.1016/j.nuclphysb.2012.10.003}},
\href{http://www.arXiv.org/abs/1207.1303}{\texttt{arXiv:1207.1303}}.
%%CITATION = ARXIV:1207.1303;%%.

\bibitem{CP5tune_CMS}
\hrefCMSnoop {}{{CMS Collaboration}, ``{Extraction and validation of a new set
  of CMS PYTHIA8 tunes from underlying-event measurements}'',} \textit{ Eur.
  Phys. J. C} \textbf{ 80} (2020) 4,
  \href{http://dx.doi.org/10.1140/epjc/s10052-019-7499-4}{\doi{10.1140/epjc/s10052-019-7499-4}},
  \href{http://www.arXiv.org/abs/1903.12179}{\texttt{arXiv:1903.12179}}.

\bibitem{Bellm:2015jjp}
\hrefCMSnoop {}{J.~Bellm {et~al.}, ``{Herwig 7.0/Herwig++ 3.0 release note}'',}
  \textit{ Eur. Phys. J. C} \textbf{ 76} (2016) 196,
  \href{http://dx.doi.org/10.1140/epjc/s10052-016-4018-8}{\doi{10.1140/epjc/s10052-016-4018-8}},
  \href{http://www.arXiv.org/abs/1512.01178}{\texttt{arXiv:1512.01178}}.

\bibitem{Pumplin:2002vw}
J.~Pumplin\hrefCMSnoop {}{ {et~al.}, ``{New generation of parton distributions
  with uncertainties from global QCD analysis}'',} \textit{ JHEP} \textbf{ 07}
  (2002) 012,
  \href{http://dx.doi.org/10.1088/1126-6708/2002/07/012}{\doi{10.1088/1126-6708/2002/07/012}},
\href{http://www.arXiv.org/abs/hep-ph/0201195}{\texttt{arXiv:hep-ph/0201195}}.
%%CITATION = HEP-PH/0201195;%%.

\bibitem{Ball:2017nwa}
\hrefCMSnoop {}{{NNPDF} Collaboration, ``{Parton distributions from
  high-precision collider data}'',} \textit{ Eur. Phys. J. C} \textbf{ 77}
  (2017) 663,
  \href{http://dx.doi.org/10.1140/epjc/s10052-017-5199-5}{\doi{10.1140/epjc/s10052-017-5199-5}},
\href{http://www.arXiv.org/abs/1706.00428}{\texttt{arXiv:1706.00428}}.
%%CITATION = ARXIV:1706.00428;%%.

\bibitem{CMS:2020dqt}
\hrefCMSnoop {}{{CMS Collaboration}, ``{Development and validation of {HERWIG}
  7 tunes from CMS underlying-event measurements}'',} \textit{ Eur. Phys. J. C}
  \textbf{ 81} (2021) 312,
  \href{http://dx.doi.org/10.1140/epjc/s10052-021-08949-5}{\doi{10.1140/epjc/s10052-021-08949-5}},
  \href{http://www.arXiv.org/abs/2011.03422}{\texttt{arXiv:2011.03422}}.

\bibitem{Martin:2010db}
\hrefCMSnoop {}{A.~D. Martin, W.~J. Stirling, R.~S. Thorne, and G.~Watt,
  ``{Heavy-quark mass dependence in global PDF analyses and 3- and 4-flavour
  parton distributions}'',} \textit{ Eur. Phys. J. C} \textbf{ 70} (2010) 51,
  \href{http://dx.doi.org/10.1140/epjc/s10052-010-1462-8}{\doi{10.1140/epjc/s10052-010-1462-8}},
  \href{http://www.arXiv.org/abs/1007.2624}{\texttt{arXiv:1007.2624}}.

\bibitem{MadGraph5_aMCatNLO}
J.~Alwall\hrefCMSnoop {}{ {et~al.}, ``{The automated computation of tree-level
  and next-to-leading order differential cross sections, and their matching to
  parton shower simulations}'',} \textit{ JHEP} \textbf{ 07} (2014) 079,
  \href{http://dx.doi.org/10.1007/JHEP07(2014)079}{\doi{10.1007/JHEP07(2014)079}},
\href{http://www.arXiv.org/abs/1405.0301}{\texttt{arXiv:1405.0301}}.
%%CITATION = ARXIV:1405.0301;%%.

\bibitem{Frederix:2012ps}
\hrefCMSnoop {}{R.~Frederix and S.~Frixione, ``{Merging meets matching in
  MC@NLO}'',} \textit{ JHEP} \textbf{ 12} (2012) 061,
  \href{http://dx.doi.org/10.1007/JHEP12(2012)061}{\doi{10.1007/JHEP12(2012)061}},
\href{http://www.arXiv.org/abs/1209.6215}{\texttt{arXiv:1209.6215}}.
%%CITATION = ARXIV:1209.6215;%%.

\bibitem{Nason:2004rx}
\hrefCMSnoop {}{P.~Nason, ``{A new method for combining NLO QCD with shower
  Monte Carlo algorithms}'',} \textit{ JHEP} \textbf{ 11} (2004) 040,
  \href{http://dx.doi.org/10.1088/1126-6708/2004/11/040}{\doi{10.1088/1126-6708/2004/11/040}},
\href{http://www.arXiv.org/abs/hep-ph/0409146}{\texttt{arXiv:hep-ph/0409146}}.
%%CITATION = HEP-PH/0409146;%%.

\bibitem{POWHEG2}
\hrefCMSnoop {}{S.~Frixione, P.~Nason, and C.~Oleari, ``{Matching NLO QCD
  computations with parton shower simulations: the POWHEG method}'',} \textit{
  JHEP} \textbf{ 11} (2007) 070,
  \href{http://dx.doi.org/10.1088/1126-6708/2007/11/070}{\doi{10.1088/1126-6708/2007/11/070}},
\href{http://www.arXiv.org/abs/0709.2092}{\texttt{arXiv:0709.2092}}.
%%CITATION = ARXIV:0709.2092;%%.

\bibitem{POWHEG3}
\hrefCMSnoop {}{S.~Alioli, P.~Nason, C.~Oleari, and E.~Re, ``{A general
  framework for implementing NLO calculations in shower Monte Carlo programs:
  the POWHEG BOX}'',} \textit{ JHEP} \textbf{ 06} (2010) 043,
  \href{http://dx.doi.org/10.1007/JHEP06(2010)043}{\doi{10.1007/JHEP06(2010)043}},
\href{http://www.arXiv.org/abs/1002.2581}{\texttt{arXiv:1002.2581}}.
%%CITATION = ARXIV:1002.2581;%%.

\bibitem{Ball:2013hta}
\hrefCMSnoop {}{{NNPDF} Collaboration, ``{Parton distributions with QED
  corrections}'',} \textit{ Nucl. Phys. B} \textbf{ 877} (2013) 290,
  \href{http://dx.doi.org/10.1016/j.nuclphysb.2013.10.010}{\doi{10.1016/j.nuclphysb.2013.10.010}},
\href{http://www.arXiv.org/abs/1308.0598}{\texttt{arXiv:1308.0598}}.
%%CITATION = ARXIV:1308.0598;%%.

\bibitem{Ball:2011uy}
\hrefCMSnoop {}{{NNPDF} Collaboration, ``{Unbiased global determination of
  parton distributions and their uncertainties at NNLO and at LO}'',} \textit{
  Nucl. Phys. B} \textbf{ 855} (2012) 153,
  \href{http://dx.doi.org/10.1016/j.nuclphysb.2011.09.024}{\doi{10.1016/j.nuclphysb.2011.09.024}},
\href{http://www.arXiv.org/abs/1107.2652}{\texttt{arXiv:1107.2652}}.
%%CITATION = ARXIV:1107.2652;%%.

\bibitem{Geant}
\hrefCMSnoop {}{{GEANT4} Collaboration, ``{\GEANTfour} --- a simulation
  toolkit'',} \textit{ Nucl. Instrum. Meth. A} \textbf{ 506} (2003) 250,
  \href{http://dx.doi.org/10.1016/S0168-9002(03)01368-8}{\doi{10.1016/S0168-9002(03)01368-8}}.

\bibitem{Sirunyan:2018shy}
\hrefCMSnoop {}{{CMS Collaboration}, ``{Evidence for associated production of a
  Higgs boson with a top quark pair in final states with electrons, muons, and
  hadronically decaying $\tau$ leptons at $\sqrt{s} =$ 13 TeV}'',} \textit{
  JHEP} \textbf{ 08} (2018) 066,
  \href{http://dx.doi.org/10.1007/JHEP08(2018)066}{\doi{10.1007/JHEP08(2018)066}},
\href{http://www.arXiv.org/abs/1803.05485}{\texttt{arXiv:1803.05485}}.
%%CITATION = ARXIV:1803.05485;%%.

\bibitem{Hocker:2007ht}
\hrefCMSnoop {}{H.~Voss, A.~H{\"o}cker, J.~Stelzer, and F.~Tegenfeldt,
  ``{TMVA}, the toolkit for multivariate data analysis with {ROOT}'',} in
  \textit{ XIth International Workshop on Advanced Computing and Analysis
  Techniques in Physics Research (ACAT)}, p.~40.
\newblock 2007.
\newblock
  \href{http://www.arXiv.org/abs/physics/0703039}{\texttt{arXiv:physics/0703039}}.
\newblock {[PoS(ACAT)040]}.
\href{http://dx.doi.org/10.22323/1.050.0040}{\doi{10.22323/1.050.0040}}.
%%CITATION = PHYSICS/0703039;%%.

\bibitem{mt21}
\hrefCMSnoop {}{C.~G. Lester and D.~J. Summers, ``{Measuring masses of
  semi-invisibly decaying particles pair produced at hadron colliders}'',}
  \textit{ Phys. Lett. B} \textbf{ 463} (1999) 99,
  \href{http://dx.doi.org/10.1016/S0370-2693(99)00945-4}{\doi{10.1016/S0370-2693(99)00945-4}},
\href{http://www.arXiv.org/abs/hep-ph/9906349}{\texttt{arXiv:hep-ph/9906349}}.
%%CITATION = HEP-PH/9906349;%%.

\bibitem{mt22}
\hrefCMSnoop {}{A.~Barr, C.~Lester, and P.~Stephens, ``{A variable for
  measuring masses at hadron colliders when missing energy is expected; m$_{\rm
  T2}$: the truth behind the glamour}'',} \textit{ J. Phys. G} \textbf{ 29}
  (2003) 2343,
  \href{http://dx.doi.org/10.1088/0954-3899/29/10/304}{\doi{10.1088/0954-3899/29/10/304}},
\href{http://www.arXiv.org/abs/hep-ph/0304226}{\texttt{arXiv:hep-ph/0304226}}.
%%CITATION = HEP-PH/0304226;%%.

\bibitem{CMS:2017sdi}
\hrefCMSnoop {}{{CMS Collaboration}, ``{Precision luminosity measurement in
  proton-proton collisions at $\sqrt{s} =$ 13 TeV in 2015 and 2016 at CMS}'',}
  \textit{ Eur. Phys. J. C} \textbf{ 81} (2021) 800,
  \href{http://dx.doi.org/10.1140/epjc/s10052-021-09538-2}{\doi{10.1140/epjc/s10052-021-09538-2}},
  \href{http://www.arXiv.org/abs/2104.01927}{\texttt{arXiv:2104.01927}}.

\bibitem{CMS:2018elu}
\href {{http://cds.cern.ch/record/2621960}}{{CMS Collaboration}, ``{CMS}
  luminosity measurement for the 2017 data-taking period at {$\sqrt{s} =
  13\TeV$}'',} {CMS Physics Analysis Summary CMS-PAS-LUM-17-004}, CERN, 2018.

\bibitem{CMS:2019jhq}
\href {{http://cds.cern.ch/record/2676164}}{{CMS Collaboration}, ``{CMS}
  luminosity measurement for the 2018 data-taking period at {$\sqrt{s} =
  13\TeV$}'',} {CMS Physics Analysis Summary CMS-PAS-LUM-18-002}, CERN, 2019.

\bibitem{CMS:2018mlc}
\hrefCMSnoop {}{{CMS Collaboration}, ``{Measurement of the inelastic
  proton-proton cross section at $ \sqrt{s}=13 $ TeV}'',} \textit{ JHEP}
  \textbf{ 07} (2018) 161,
  \href{http://dx.doi.org/10.1007/JHEP07(2018)161}{\doi{10.1007/JHEP07(2018)161}},
  \href{http://www.arXiv.org/abs/1802.02613}{\texttt{arXiv:1802.02613}}.

\bibitem{CMS:2021icx}
\hrefCMSnoop {}{{CMS Collaboration}, ``{Measurement of the inclusive and
  differential WZ production cross sections, polarization angles, and triple
  gauge couplings in pp collisions at $\sqrt{s}$ = 13 TeV}'',} 2021.
  \href{http://www.arXiv.org/abs/2110.11231}{\texttt{arXiv:2110.11231}}.
  Submitted to \textit{JHEP}.

\bibitem{Ball:2014uwa}
\hrefCMSnoop {}{{NNPDF} Collaboration, ``Parton distributions for the {LHC Run
  II}'',} \textit{ JHEP} \textbf{ 04} (2015) 040,
  \href{http://dx.doi.org/10.1007/JHEP04(2015)040}{\doi{10.1007/JHEP04(2015)040}},
\href{http://www.arXiv.org/abs/1410.8849}{\texttt{arXiv:1410.8849}}.
%%CITATION = ARXIV:1410.8849;%%.

\bibitem{Butterworth:2015oua}
\hrefCMSnoop {}{J.~Butterworth {et~al.}, ``{PDF4LHC recommendations for LHC Run
  II}'',} \textit{ J. Phys. G} \textbf{ 43} (2016) 023001,
  \href{http://dx.doi.org/10.1088/0954-3899/43/2/023001}{\doi{10.1088/0954-3899/43/2/023001}},
  \href{http://www.arXiv.org/abs/1510.03865}{\texttt{arXiv:1510.03865}}.

\bibitem{Barlow:1993dm}
\hrefCMSnoop {}{R.~J. Barlow and C.~Beeston, ``{Fitting using finite Monte
  Carlo samples}'',} \textit{ Comput. Phys. Commun.} \textbf{ 77} (1993) 219,
\href{http://dx.doi.org/10.1016/0010-4655(93)90005-W}{\doi{10.1016/0010-4655(93)90005-W}}.
%%CITATION = CPHCB,77,219;%%.

\bibitem{Conway:2011in}
\hrefCMSnoop {}{J.~S. Conway, ``{Incorporating nuisance parameters in
  likelihoods for multisource spectra}'',} in \textit{ {PHYSTAT 2011}}, p.~115.
\newblock 2011.
\newblock \href{http://www.arXiv.org/abs/1103.0354}{\texttt{arXiv:1103.0354}}.
\newblock
  \href{http://dx.doi.org/10.5170/CERN-2011-006.115}{\doi{10.5170/CERN-2011-006.115}}.

\bibitem{Cowan:2010js}
\hrefCMSnoop {}{G.~Cowan, K.~Cranmer, E.~Gross, and O.~Vitells, ``{Asymptotic
  formulae for likelihood-based tests of new physics}'',} \textit{ Eur. Phys.
  J. C} \textbf{ 71} (2011) 1554,
  \href{http://dx.doi.org/10.1140/epjc/s10052-011-1554-0}{\doi{10.1140/epjc/s10052-011-1554-0}},
  \href{http://www.arXiv.org/abs/1007.1727}{\texttt{arXiv:1007.1727}}.
[Erratum: \DOI{10.1140/epjc/s10052-013-2501-z}].
%%CITATION = ARXIV:1007.1727;%%.

\bibitem{CMS-NOTE-2011-005}
\href {https://cds.cern.ch/record/1379837}{{ATLAS and CMS Collaborations, LHC
  Higgs Combination Group}, ``{Procedure for the LHC Higgs boson search
  combination in summer 2011}'',} Technical Report CMS-NOTE-2011-005.
  ATL-PHYS-PUB-2011-11, CERN, 2011.

\bibitem{Khachatryan:2014jba}
\hrefCMSnoop {}{{CMS Collaboration}, ``Precise determination of the mass of the
  {Higgs} boson and tests of compatibility of its couplings with the standard
  model predictions using proton collisions at 7 and 8 {TeV}'',} \textit{ Eur.
  Phys. J. C} \textbf{ 75} (2015) 212,
  \href{http://dx.doi.org/10.1140/epjc/s10052-015-3351-7}{\doi{10.1140/epjc/s10052-015-3351-7}},
\href{http://www.arXiv.org/abs/1412.8662}{\texttt{arXiv:1412.8662}}.
%%CITATION = ARXIV:1412.8662;%%.

\bibitem{supp}
\hrefCMSnoop {}{}See the Supplemental Material at [URL will be inserted by
  publisher] for distributions of the kinematic variables used for the training
  of BDT discriminants and the two BDT discriminants.

\bibitem{Anastasiou:2003ds}
\hrefCMSnoop {}{C.~Anastasiou, L.~J. Dixon, K.~Melnikov, and F.~Petriello,
  ``{High precision QCD at hadron colliders: electroweak gauge boson rapidity
  distributions at NNLO}'',} \textit{ Phys. Rev. D} \textbf{ 69} (2004) 094008,
  \href{http://dx.doi.org/10.1103/PhysRevD.69.094008}{\doi{10.1103/PhysRevD.69.094008}},
\href{http://www.arXiv.org/abs/hep-ph/0312266}{\texttt{arXiv:hep-ph/0312266}}.
%%CITATION = HEP-PH/0312266;%%.

\bibitem{fewz2p1}
\hrefCMSnoop {}{R.~Gavin, Y.~Li, F.~Petriello, and S.~Quackenbush, ``{W physics
  at the LHC with FEWZ 2.1}'',} \textit{ Comput. Phys. Commun.} \textbf{ 184}
  (2013) 208,
  \href{http://dx.doi.org/10.1016/j.cpc.2012.09.005}{\doi{10.1016/j.cpc.2012.09.005}},
\href{http://www.arXiv.org/abs/1201.5896}{\texttt{arXiv:1201.5896}}.
%%CITATION = ARXIV:1201.5896;%%.

\bibitem{HEPData}
\hrefCMSnoop {}{``{HEPD}ata record for this analysis'',} 2022.
\newblock
  \href{http://dx.doi.org/10.17182/hepdata.130562}{\doi{10.17182/hepdata.130562}}.

\end{thebibliography}\endgroup
